\documentclass[pre,aps,twocolumn,nopacs,floatfix,superscriptaddress]{revtex4-1}

\usepackage{amsfonts,amssymb,amsmath,array}
\usepackage{graphicx}
\usepackage{amsbsy}
\usepackage{color}
\usepackage{hyperref} 
\usepackage{enumerate}
\usepackage{mathtools}
\usepackage{lipsum} 
\usepackage{cleveref}                     
\begin{document}

\title{Thermodynamic bounds for diffusion\\ in non-equilibrium systems with multiple timescales}

\author{A. Plati}
\affiliation{Department of Physics, University of Rome Sapienza, P.le Aldo Moro 2, 00185, Rome, Italy}
\affiliation{Institute for Complex Systems - CNR, P.le Aldo Moro 2, 00185, Rome, Italy}
\affiliation{Universit\'e Paris-Saclay, CNRS, Laboratoire de Physique des Solides, 91405 Orsay, France}
\author{A. Puglisi}
\affiliation{Department of Physics, University of Rome Sapienza, P.le Aldo Moro 2, 00185, Rome, Italy}\affiliation{Institute for Complex Systems - CNR, P.le Aldo Moro 2, 00185, Rome, Italy}
\affiliation{INFN, University of Rome Tor Vergata, Via della Ricerca Scientifica 1, 00133, Rome, Italy}

\author{A. Sarracino}
\affiliation{Department of Engineering, University of Campania ``Luigi Vanvitelli'', 81031 Aversa (CE), Italy}
\affiliation{Institute for Complex Systems - CNR, P.le Aldo Moro 2, 00185, Rome, Italy}
              
\begin{abstract}

We derive a Thermodynamic Uncertainty Relation bounding the mean
squared displacement of a Gaussian process with memory, driven out of
equilibrium by unbalanced thermal baths and/or by external forces. Our
bound is tighter with respect to previous results and also holds at
finite time.  We apply our findings to experimental and numerical data
for a vibro-fluidized granular medium, characterized by regimes
of anomalous diffusion. In some cases, our relation can distinguish
between equilibrium and non-equilibrium behavior, a non-trivial
inference task, particularly for Gaussian processes.
  
\end{abstract}

\maketitle

\section{Introduction}

The relation between dynamical properties of a
system and its thermodynamics plays a central role in modern
non-equilibrium statistical physics.  In systems composed by many
interacting particles, it is common to observe different phenomena
occurring at different timescales, the paradigmatic example being the
several regimes of structural relaxation in undercooled
liquids~\cite{cavagna2009supercooled}. This complex dynamics usually
gives place to a mean squared displacement (MSD) of some fluctuating
observable which shows several non-diffusive regimes.  For
instance, the diffusion of particles in liquids often displays
transient sub-diffusive or flat MSD corresponding to cage
effects. Interestingly, these regimes are also observed in liquid-like
systems realized by replacing molecules with macroscopic spheres, in
the context of dense vibro-fluidized granular materials, both in
simulations and in
experiments~\cite{marty2005subdiffusion,bodrova2012intermediate,scalliet2015cages,plati2020slow}. Additionally,
these systems can display novel phenomena such as a superdiffusive
transient regime after the cage stage and before the final asymptotic
standard diffusion~\cite{plati2019dynamical}.  While the molecular
liquid case is typically at thermal equilibrium (even if under sudden
quench the relaxation time may diverge and shift the system into
non-equilibrium), a vibrated granular medium is intrinsically out of
equilibrium, even if stationary, because of the presence of several
energy flows from and into the system (friction, inelastic collisions,
external energy pumping, etc).  In principle, however, diffusion
properties are not evidently related to the status of equilibrium or
non-equilibrium~\cite{lasanta2015itinerant}. It is therefore important
to explore the existence of physical constraints that could restrict
the possible behaviors of the MSD and relate certain observations to
the thermodynamic status of the system~\cite{seifert2019stochastic}.

Recently, an important step in building a bridge between anomalous
dynamical regimes and thermodynamic properties has been done
exploiting the Thermodynamic Uncertainty Relations
(TUR)~\cite{Barato2015,seifert2018stochastic}. These relations, valid
for quite a large class of stochastic processes, also demonstrated
through several different
routes~\cite{Gingrich2016,Hasegawa2019,Hasegawa2019II,dechant2020fluctuation},
typically take the form
\begin{equation}
\frac{\langle \Delta\theta(t)^2 \rangle}{\langle  \theta(t)
  \rangle^2} \ge \frac{2}{\langle S(t)\rangle},\label{TUR}
\end{equation}
where $\Delta\theta(t)=\theta(t) - \langle\theta(t) \rangle$ and $ \theta(t)=\int_0^tdt'~\omega(t')$ is an integrated current over a time $t$, while $S(t)$ is the
entropy produced by the system in the same time interval. Here and in the next we fix $k_b=1$. Identifying $\theta(t)$ as the displacement of a particle with velocity $\omega(t)$ and multiplying both sides of \eqref{TUR} for $\langle  \theta(t)
  \rangle^2$, one obtains a straightforward application to
the MSD, which has been applied to the case of overdamped systems
with {\em two} dynamical regimes, one being anomalous and one
being standard~\cite{Hartich2021}. In particular it has been shown
that the TUR implies a minimum (or maximum) time of validity for the
super- (or sub-) diffusion.

The application of this kind of results to non-equilibrium systems with multiple characteristic timescales requires a more general and effective bound, which is
the purpose of the present paper. Here we show how to extend TURs for
underdamped dynamics to the case of systems with multiple timescales
and multiple baths, such as active liquids and vibrofluidized granular media. We obtain a general formula to bound the MSD in
time, with the interesting and unexpected result that only a part of
the entropy production enters the bound, making it tighter than that
one would get using the whole entropy production. We present
analytical results within the framework of Markovian continuous
linear systems, that can emerge from the Markovianization of systems
with memory, representing therefore a very general tool for the study
of coarse-grained variables in presence of hydrodynamic backflow \cite{Loos2021,Franosch2011} and in out-of-equilibrium many-body
systems~\cite{zamponi2005fluctuation,puglisi2009irreversible,crisanti2012nonequilibrium},
including driven macroscopic dissipative systems such as
granular materials~\cite{puglisi2014transport} and active
matter~\cite{rizkallah2022microscopic}. We recall that general thermodynamic bounds for underdamped dynamics still represent an open problem \cite{Hasegawa2019,Fischer2020,Lee2021,Dechant2022} while a TUR for non-Markovian system has been previously derived for a very general class of memory kernels but always assuming thermal equilibrium with a single thermostat \cite{di2020thermodynamic}.

Our results are successfully applied to numerical and experimental data coming from two different systems of interacting particles where an intruder is immersed in a vibrated granular fluid \cite{scalliet2015cages,sarracino2010irreversible}.
Remarkably, our approach also shows that, in the zero driving limit,
we obtain a TUR for the spontaneous diffusion that in fact can be
tested with systems in the absence of an external bias, allowing one
to distinguish between equilibrium and non-equilibrium behavior.

\section{The model}

We consider a set of $n+1$ coupled dynamical
variables, each in contact with a different thermal bath. The first
variable represents the main observable, possibly subject to a
constant external force, while the other $n$ variables are auxiliary
variables, representing memory terms. This kind of model can describe
the underdamped dynamics of a tracer in a fluid, when a separation of timescales
allows one to obtain an effective generalized Langevin equation (GLE)
for the slow variable~\cite{cortes1985generalized}, or systems with feedback control~\cite{munakata2013feedback,munakata2014entropy,costanzo2021stochastic,Costanzo2022}. Defining the
vectors $\boldsymbol{X}=\{\omega,\Omega_1,\ldots,\Omega_n\}$,
$\boldsymbol{\xi}=\{\xi_0,\xi_1,\ldots,\xi_n\}$ and
$\boldsymbol{F}=\{F_{\text{ext}},0,\ldots,0\}$, the dynamics is
described by the coupled equations:
\begin{equation}\label{model}
\dot{\boldsymbol{X}}=\hat{A}\boldsymbol{X}+\hat{B}\boldsymbol{\xi} +\boldsymbol{F},
\end{equation}
where $\xi_i$ are uncorrelated white noises with zero mean and unit
variance while the two matrices $\hat{A}$ and $\hat{B}$ are given by:
\begin{equation}\label{eq::GeneralMatrix}
\hat{A}=\left( {\begin{array}{cccc}
    -1/\tau & 1/b_1 & \ldots & 1/b_n \\
   -a_1 b_1 & -1/\tau_1 &0 & 0\\
   \vdots& 0 & \ddots&0 \\
    -a_n b_n& 0 & 0 &-1/\tau_n \\ 
  \end{array} } \right), 
\end{equation}
\begin{equation}\label{eq::matB}
\quad \hat{B}=\text{diag}\left(\sqrt{2q/\tau},\sqrt{2q_1a_1b_1^2/\tau_1},\ldots,\sqrt{2q_na_nb_n^2/\tau_n} \right).
\end{equation}
Here the $\tau_i$s, the $b_i$s and the $a_i$s are positive parameters
with the dimension of time, time and inverse squared time,
respectively. We consider $\omega$ odd and the $\Omega_i$s even under
time reversal. With this choice the fluctuating entropy production takes the form of heat exchanges over effective temperatures (see Appendix~\ref{Sec::EntProd}). We also propose a physical interpretation of this time reversal symmetries in Appendix~\ref{subs:SymTimeAux}. We note that $\hat{A}$ is an arrowhead matrix, namely it has non-zero elements only in the first row, in the
first column and in the principal diagonal. This form has a physical
meaning: The auxiliary variables describe the memory in the system,
each one has a characteristic relaxation time $\tau_i$ and is coupled
with the main observable only. The above equations are indeed
equivalent to the following GLE~\cite{puglisi2009irreversible}:
\begin{eqnarray}
 \dot{\omega}(t)=-\int_{-\infty}^t\gamma(t-t')\omega(t')dt'+\eta_s(t) +F_{\text{ext}},\label{GLE} \\
 \gamma(t)=\frac{2}{\tau}\delta(t)+\sum_{k=1}^n a_k e^{-\frac{t}{\tau_k}},\\
 \langle \eta_s(t)\eta_s(t') \rangle=\frac{2q}{\tau}\delta(|t-t'|)+ \sum_{k=1}^n q_k a_k e^{-\frac{|t-t'|}{\tau_k}},
\end{eqnarray}
and the auxiliary variables are:
\begin{equation}
\Omega_k=-b_k\int_{-\infty}^t dt' e^{-\frac{t-t'}{\tau_k}}\left[a_k\omega(t')-\sqrt{\frac{2q_k a_k}{\tau_k}}\xi_k(t')\right]. 
\end{equation}
Interestingly, a memory kernel which is a sum of a few exponential
decays can approximate also non-exponential kernels, such as power-law
decays typical of several transport phenomena in dense
systems~\cite{min2005observation} (see also Appendix~\ref{sec:powerLaw}).
We recall here that the use of exponential memory kernels to describe the diffusion of an intruder in a complex fluid is motivated by a typical approximation done for Brownian motion at high densities when the coupling with hydrodynamic modes decaying exponentially in time is taken into account~\cite{sarracino2010irreversible,Berne66}.

We point out that this model is built in such a way to recover the
fluctuation-dissipation relation of the second kind $\langle
\eta_s(t)\eta_s(t') \rangle=q\gamma(|t-t'|)$ if all the thermostats
are at the same temperature $q_k=q$. With this condition (and $F_{\text{ext}}=0$),
thermodynamic equilibrium is properly described. In the Fokker-Planck formalism this is equivalent to a null irreversible current~\cite{Gardiner} (see also Appendix~\ref{subs:eqFPE}). The solution for the
stationary probability distribution function is a multivariate
Gaussian $P(\boldsymbol{X})\propto \exp\left(- \Delta \boldsymbol{X}
\left.\hat{\beta} \Delta\boldsymbol{X} \right/2\right)$ where $\Delta
\boldsymbol{X}=\boldsymbol{X}-\langle \boldsymbol{X} \rangle$ and
$\hat{\beta}$ is the inverse of the covariance matrix
$\sigma_{ij}=\langle \Delta X_i \Delta X_j \rangle$. Note that, thanks to the linearity of the model, $\hat{\beta}$ and $\hat{\sigma}$ do not depend on  $F_{\text{ext}}$. Such a distribution
is canonical ($\beta_{ij} \propto \delta_{ij}/q$) at
equilibrium (see Appendix~\ref{sec:statics}). 
We remark that the model has two different sources of
non-equilibrium: The coupling with different thermal baths (i.e. when $q,q_k$ are different) and the external force
$F_{\textrm{ext}}$. Interestingly, the second ingredient triggers an
average drift $\langle \boldsymbol{X} \rangle \neq 0$, while the first
one does not.

The entropy production rate (EPR)~\cite{lebowitz1999gallavotti}  of the model in the steady state
reads (see \ref{Sec::EntProd} for details):  
\begin{equation}\label{eq::EntProdTOTALgen}
  \langle \dot{S}\rangle=\langle \dot{S}\rangle_{\textrm{ext}}+\langle \dot{S}\rangle_{\textrm{th}},
\end{equation}
where we defined an external contribution due to the presence of
forcing $\langle \dot{S}\rangle_{\textrm{ext}}=\frac{1}{q}\langle \omega\rangle
F_{\text{ext}}+\sum_i(\frac{1}{q} - \frac{1}{q_i})\frac{\langle \omega\rangle
  \langle \Omega_i\rangle}{b_i}$ and one $\langle
\dot{S}\rangle_{\textrm{th}}=\sum_i \frac{1}{b_i} (\frac{1}{q} -
\frac{1}{q_i})\sigma_{0i}$ due  only to the
coupling with baths at different temperatures. This last term is
positive because it is the only contribution in the absence of the
external driving~\footnote{Note that, due to the linearity, the covariances $\hat{\sigma}_{0i}$ do not depend upon $F_{\text{ext}}$}.
 The mean values of the dynamical
variables in the steady state are:
\begin{equation}\label{mean}
  \langle \omega\rangle = \frac{F_{\textrm{ext}}\tau}{1+\tau\sum_k \tau_ka_k},  \quad
  \langle \Omega_i\rangle = -\tau_i a_i b_i \langle \omega\rangle
\end{equation}
which imply that $\langle \dot{S}\rangle_{\textrm{ext}}$ is proportional to $F_{\text{ext}}^2$.

\section{TUR in the large time limit}
We start by considering the bound for the diffusion coefficient of the tracer obtained from the TUR~\cite{Barato2015,Gingrich2016} valid for overdamped dynamics in the large time limit of the stationary state
\begin{equation}
\lim_{t\to \infty}\frac{\langle \Delta\theta(t)^2 \rangle}{t}\ge \frac{2\langle \omega \rangle^2}{\langle \dot{S}\rangle}\label{TUR_Over}
\end{equation}  
where $\Delta\theta(t)$ is defined as in Eq.~\eqref{TUR}. 
In our model, all the terms of the above inequality can be
explicitly computed.  Indeed, we can relate the spectrum and the diffusion coefficient with the Wiener-Khinchin theorem
$\mathcal{S}_{00}(0)=\lim_{t\to
  \infty}\langle\Delta\theta(t)^2\rangle/t$, where the spectral matrix
is defined as the Fourier transform of the stationary correlation matrix:  
\begin{equation}
\hat{\mathcal{S}}(f)\equiv \int_{-\infty}^{+\infty}dt e^{-if t}\hat{\sigma}(t)=(\hat{A}+i\hat{I} f)^{-1}\hat{B}\hat{B}^T(\hat{A}^T-i\hat{I}
f)^{-1}
\end{equation}
where $\sigma_{ij}(t-s)=\langle \Delta X_i (t) \Delta X_j (s) \rangle$ and $\hat{I}$ is the identity matrix. Inverting the
arrowhead matrix $\hat{A}$~\cite{Wani2016}, we get (see Appendix~\ref{sec::diffcoeff})
\begin{equation}\label{eq::generalDiffCoeff}
\mathcal{S}_{00}(0)=\left[\hat A^{-1}\hat B\hat B^T(\hat A^T)^{-1}\right]_{00}=\mathcal{D}_{\text{eq}}\left[ \frac{1+\tau\sum_k \frac{q_k}{q}a_k\tau_k}{1+\tau\sum_k a_k \tau_k} \right]
\end{equation}
where $\mathcal{D}_{\text{eq}}=2q\tau/(1+\tau\sum_k a_k\tau_k)$ is the diffusion coefficient when $q_i=q$ $\forall$ $i$. Then, using Eqs.~\eqref{eq::EntProdTOTALgen} and~(\ref{mean}), we have
\begin{equation}\label{eq::generalBoundAsint}
\frac{2\langle \omega \rangle^2}{\langle \dot{S}\rangle_{\text{ext}}}=\mathcal{D}_{\text{eq}}\left[\frac{1+\tau\sum_k a_k \tau_k}{1+\tau\sum_k \frac{q}{q_k}a_k\tau_k} \right].
\end{equation}
From this expression we arrive to the following relation (see Appendix~\ref{sec::turlargetimeapp} for details):
\begin{equation}\label{eq:TurOverBetter}
\lim_{t\to \infty}\frac{\langle \Delta\theta(t)^2 \rangle}{t}\ge \frac{2\langle \omega \rangle^2}{\langle \dot{S}\rangle_{\text{ext}}}\ge \frac{2\langle \omega \rangle^2}{\langle \dot{S}\rangle}.
\end{equation}
This shows that in our model a bound tighter than the one of
Eq.~(\ref{TUR_Over}) can be obtained, by considering in the EPR  the contribution $\langle
\dot{S}\rangle_{\text{ext}}$ only.  Below we extend this result to finite times.

As an additional remark, we note that completely ignoring the presence of
thermostats with different temperatures can imply a violation of the associated
inequality. Indeed, defining the contribution associated with
the drift $\langle \dot{S} \rangle_{\text{drift}}=\langle \omega \rangle
F_{\text{ext}}/q$, one can verify that the inequality $\mathcal{S}_{00}(0)
\ge 2\langle \omega \rangle^2/\langle \dot{S} \rangle_{\text{drift}}=\mathcal{D}_{\text{eq}}$
is violated if $\sum_k (q_k-q)a_k\tau_k < 0$.

\section{TUR at finite times}

To derive the general finite-times
expression of a TUR with a tighter bound, we can proceed as
in~\cite{Hasegawa2019II}.  We consider a fictive $h$-dynamics
(generating $\langle\cdots\rangle_h$ averages over a distribution
$P_h$) that coincides with the original one as $h=0$, and write the
Cram\'er-Rao inequality for an unbiased estimator $\Theta$ of a
function $\psi(h)$:
\begin{equation}\label{eq::CramRao}
 \frac{\text{Var}_h(\Theta[\Gamma_t])}{[\partial_h \langle \Theta[\Gamma_t] \rangle_{h}]^2} \ge \frac{1}{\mathcal{I}_{\text{F}}(h)}.
 \end{equation} 
Here $\Gamma_t$ is the stochastic trajectory of duration $t$ along
which the estimator is evaluated and $\mathcal{I}_{\text{F}}$ is the
Fisher information~\cite{cover1999elements}. Thus we have $\langle
\Theta[\Gamma_t] \rangle_{h}=\psi(h)$ and we require that
$(\partial_h \langle \Theta[\Gamma_t]
\rangle_{h})|_{h=0}=\langle \Theta[\Gamma_t] \rangle$, so
that the lhs of Eq.~(\ref{eq::CramRao}) calculated in $h=0$
coincides with the uncertainty of the generalized current $\langle
\Theta[\Gamma_t] \rangle$. Note that this condition depends both on
how the current is defined and on the choice of the fictive
dynamics~\citep{Hasegawa2019,Lee2021}.

To derive the TUR with the tighter bound, we introduce a perturbation
to Eq.~(\ref{model}) in the form $h \boldsymbol{V}$, where
$\boldsymbol{V}=\{\langle \omega \rangle/\tau,-\langle \Omega_1
\rangle/\tau_1,\ldots,-\langle \Omega_n \rangle/\tau_n\}$.  With this
choice, evaluating the Cram\'er-Rao inequality for $h=0$ in the
stationary state, we get (see Appendix~\ref{secc::cramerraoapp} for details)
\begin{equation}
\frac{\langle \Delta\theta(t)^2 \rangle}{(\langle \omega \rangle t)^2} \ge \frac{2}{\Delta S_{ext}(t)+\mathcal{I}},
\end{equation}
where   
\begin{subequations}
\begin{equation}
  \mathcal{I}=2\int d\boldsymbol{X} \frac{[\partial_h P_h(\boldsymbol{X})]^2|_{h=0}}{P( \boldsymbol{X})}, 
\end{equation}
\begin{equation}
  \Delta S_{\text{ext}}(t)= \int_0^t dt' \left[ \frac {\langle \omega \rangle^2}{\tau q} + \sum_i \frac{\langle \Omega_i \rangle^2}{\tau _i q_i a_i b_i^2} \right] 
  =\langle \dot{S} \rangle_{\text{ext}}t.
  \end{equation}
\end{subequations}
The above expression coincides with the definition of $\langle \dot{S}
\rangle_{\text{ext}}$ below Eq.~\eqref{eq::EntProdTOTALgen} (see Appeendix~\ref{sec::crdetails}).  We then obtain the following TUR for the MSD also valid
at finite times in the steady state, that is consistent with the improved bound discussed
for large times, Eq.~(\ref{eq:TurOverBetter})
\begin{equation}\label{eq::finalBound}
\langle \Delta\theta(t)^2 \rangle\ge \frac{2\langle \omega \rangle^2 t^2}{\langle \dot{S} \rangle_{\text{ext}}t+\mathcal{I} }.  
\end{equation}
Exploiting the linearity of the model we can easily obtain $P_h$ from which we compute the explicit form of the non-extensive term: 
\begin{equation}\label{eq::nonExt}
\mathcal{I}=2\langle \omega \rangle^2 \beta_{00}.
\end{equation}
It is important to note that $\langle \omega \rangle^2$ simplifies in
the rhs of the TUR \eqref{eq::finalBound}, making it independent of
$F_{\text{ext}}$, as the lhs. Thus, for $F_{\text{ext}} \to 0$, the
bound remains finite, at variance with the weaker bound obtained from
the total EPR $\langle \dot{S} \rangle$. Eq.~\eqref{eq::finalBound} therefore also works in the case of
force-free diffusion, as shown in the following. 
We remark that even if the model is linear, an analytical form for the MSD when $n>1$ can be quite involved~\cite{Loos2021}. A bound with a simple functional form as the one provided by formula~\eqref{eq::finalBound} can be, therefore, precious.
It is interesting to consider also the consequence of some lack of information in the
modeling procedure: for instance one could overlook the different
thermostats, and could be tempted to use the asymptotic bound considering just $\langle \dot{S} \rangle_{\text{drift}}$ for the whole available time-range
(which is appealing as it is simpler and does not require estimating
$\mathcal{I}$). We denote this case as the ``incomplete bound'' (IB) and discuss its consequences in the following examples.

\section{Tracer dynamics in a dense granular medium}
In order to illustrate the validity of our results and to show their
relevance in physical systems, we apply them in the case of diffusion
in driven granular fluids. We consider the case $n=1$, that has been
shown to describe the behavior of a massive tracer in a moderately
dense granular medium ~\cite{sarracino2010irreversible}. In this
conditions the MSD of the tracer can exhibit a subdiffusive behavior
at intermediate times due to the caging effect of the surrounding
grains. We compare the bound \eqref{eq::finalBound} with MSD of this
kind obtained in experiments \cite{scalliet2015cages} and molecular
dynamics simulations \cite{sarracino2010irreversible}. In the
experiment, the tracer diffuses in a system of steel spheres confined
in a 3D box vertically driven by an electrodynamic shaker, while
numerical simulations consider the 2D case of hard dissipative disks coupled to
a spatially homogeneous thermostat.
We use the two-dimensional form of Eq.~\eqref{model} with $a_1=\frac{\alpha}{\tau\tau_1}$ and $b_1=\tau$ obtaining the same model used in~\cite{sarracino2010irreversible}.
The mean values $\langle \omega \rangle=\tau F_{\text{ext}}/(1+\alpha)$
and $\langle \Omega_1 \rangle=-\alpha\langle \omega
\rangle$ appear in
the EPR:
\begin{equation}\label{eq:EntProdDrNonEq}
\langle \dot{S}\rangle = \frac{1}{q}\langle \omega\rangle F_{\text{ext}} -
\left[ \frac{q_1-q}{\tau q q_1} \right]\alpha \langle \omega
\rangle^2 + \left[ \frac{q_1-q}{\tau q q_1} \right]
\hat{\sigma}_{01}.
\end{equation}

\begin{figure}
\centering
\includegraphics[width=\columnwidth,clip=true]{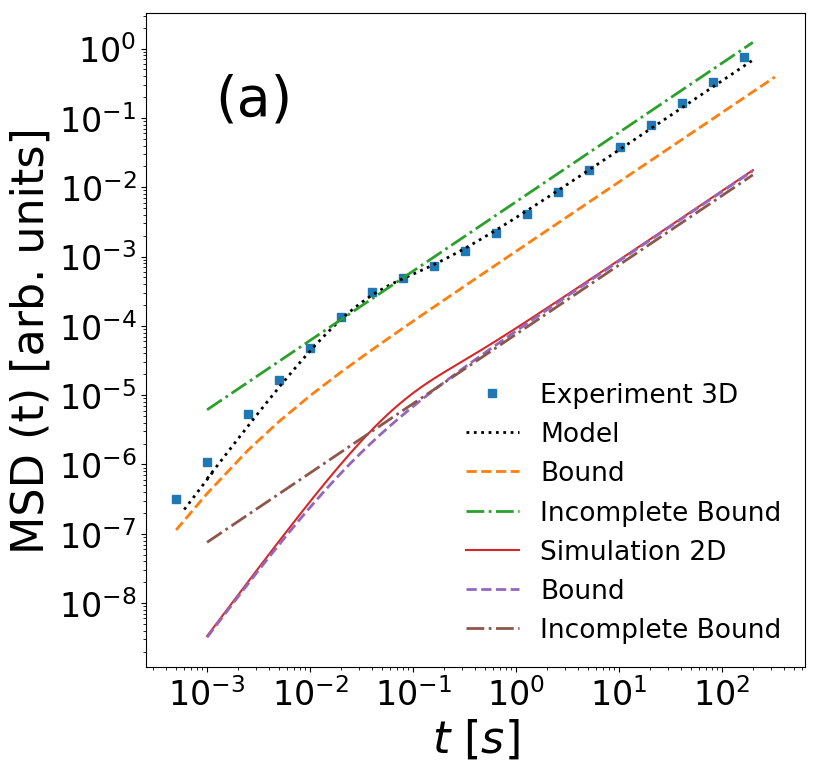}
\includegraphics[width=\columnwidth,clip=true]{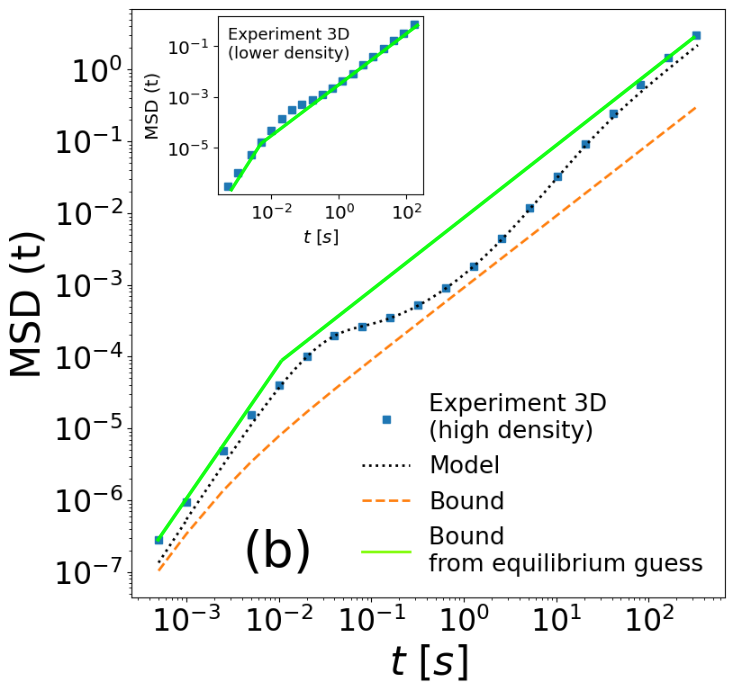}
\caption{MSD of a large intruder immersed in a vibrated granular fluid at moderate density (a) and at high density (b). In both cases the bounds are calculated with the numerical values of the model parameters obtained by fitting the data. Details on the fitting procedure are given in Appendix~\ref{sec::Fit}. In panel b, the equilibrium guess is constructed by connecting the two slopes of the ballistic and the diffusive regime following what we would expect at equilibrium from Eq.~\eqref{eq::BoundEqLimit}. The inset shows an MSD (same experimental data of panel a) whose form is compatible with thermodynamic equilibrium. \label{fig:Fig1}}
\end{figure}
The comparison of the bounds discussed above with the MSD measured in
experiments and simulations is shown in Fig.~\ref{fig:Fig1}(a).  Here we
see that the bound from Eq.~\eqref{eq::finalBound} (dashed lines) is
close (from below) to the data at all timescales. The IB (dot-dashed
lines) is obtained neglecting the different thermostats. Two possible
situations may appear: i) the IB is valid at late times but (as
expected) violated at short times (see curves for the 2D simulations),
ii) it is violated also in the diffusive regime (see data for the 3D
experiment). The difference between these two conditions depends on
the interplay of characteristic times and temperatures. Since data come from force-free diffusion, we used the bound in the limit $F_{\text{ext}} \to 0$, which is meaningful for Eq.~\eqref{eq::finalBound} while trivial for Eq.~\eqref{TUR_Over}. This is the reason why we don't compare our bound with the one obtained from the standard TUR in Fig.~\ref{fig:Fig1}(a). 

The bound on the extent of non-diffusive regimes of our model is
discussed in Appendix~\ref{sec::boundonext}. We stress that a valid
TUR without the hypothesis of velocity relaxation is necessary in this
class of model because, contrary to what happens in
\cite{Hartich2021}, the MSD predicted by our model always exhibits a
ballistic regime at short times. Then, in order to correctly bound the
extent of anomalous diffusion, we need a thermodynamic bound that is
not simply linear in time.

\section{Forbidden equilibrium regimes}

The derived bound Eq.~\eqref{eq::finalBound} holds on a class of models for which the analytical expression of many thermodynamic quantities is available \cite{Loos2021,Gardiner}. Thus, it is important to specify for which  practical purpose one can exploit our bound. In view of this, here we show how Eq.~\eqref{eq::finalBound} can be used to directly infer non-equilibrium signatures from data. We consider the rhs of Eq.~\eqref{eq::finalBound} at equilibrium and we refer to it as
$\mathcal{B}_{\text{eq}}(t)$. We equate all the thermostats in
Eqs.~\eqref{eq::generalBoundAsint} and~\eqref{eq::generalDiffCoeff}
and take $\hat{\sigma}$ diagonal in Eq.~\eqref{eq::nonExt}, obtaining
\begin{equation}\label{eq::BoundEqLimit}
\mathcal{B}_{\text{eq}}(t) \sim \langle \Delta\theta(t)^2 \rangle \sim \begin{cases} \langle\Delta \omega^2 \rangle t^2 \quad t \ll \mathcal{I}/\langle \dot{S} \rangle_{\text{ext}} \\ \mathcal{D}_{\text{eq}} t \quad t \gg \mathcal{I}/\langle \dot{S} \rangle_{\text{ext}}. \end{cases}  
\end{equation} 
Note that $\mathcal{I}/\langle \dot{S}\rangle$ is always well defined
at equilibrium since the quadratic dependence on $F_{\text{ext}}$ cancels out.
Provided that the system is in equilibrium,
Eq.~\eqref{eq::BoundEqLimit} shows that the MSD and the bound coincide
in both the short and long time limit while for intermediate times the
inequality holds. This observation allows one to exclude the
occurrence of certain transient anomalous diffusion regimes at
equilibrium or, equivalently, to ensure that certain forms of MSD are
compatible only with non-equilibrium dynamics. This test for
equilibrium compatibility can be done by connecting the two slopes of
the ballistic and diffusive regimes of a given MSD in a log-log plot
and considering this curve as a lower bound from an equilibrium
guess. Indeed, given the functional form of the bound
Eq.~\eqref{eq::finalBound} and knowing that it reduces to an equality
at short and long times if $q_i=q$ $\forall$ $i$
(Eq.~\eqref{eq::BoundEqLimit}), an MSD coming from an equilibrium
dynamics is expected to lie above the constructed curve at all
times. Then, when any tract of the MSD is found to lie below the lower
bound from the equilibrium guess, then one can deduce that, if the
dynamics follows Eq.~\eqref{model}, \eqref{eq::GeneralMatrix} and \eqref{eq::matB}, the observed MSD is not
compatible with thermodynamic equilibrium.  To illustrate this
application, we consider the case $n=2$, that can describe the
anomalous diffusion of a tracer in a dense granular system with very
slow characteristic times.  We take $a_1=\frac{\alpha}{\tau\tau_1}$,
$a_2=\frac{\epsilon^2}{\tau\tau_2}$ and $b_1=b_2=\tau$, where
$\epsilon=\tau/\tau_2$. For $\epsilon \to 0$ and keeping finite the
amplitude of the noise $\xi_2$, we obtain the same model described
in~\cite{plati2020slow}. As we can see in Fig.~\ref{fig:Fig1}(b), this
model can properly reproduce the experimental data of the MSD,
characterized by a surprising superdiffusive regime after the cage
subdiffusion. Its origin relies on the presence of a slow collective
motion of the granular medium due to the interplay of disorder and
friction~\cite{plati2019dynamical,plati2022friction}. As evident from
Fig.~\ref{fig:Fig1}(b), the behavior of the MSD is not compatible with
the bound guessed from the equilibrium
condition~\eqref{eq::BoundEqLimit}. Then, we can conclude that the
underlying dynamics is out of equilibrium without performing any
further analysis. In order to complete the picture, we show in the
inset of Fig.~\ref{fig:Fig1}(b) the application of this procedure to
the experimental data of Fig.~\ref{fig:Fig1}(a) which come from a less
dense system where the slow collective motion and the consequent
superdiffusive regime do not appear. In this case, the MSD lies always
above the equilibrium guess so we cannot draw any conclusion on the
non-equilibrium properties of the dynamics without estimating the
model's parameters.

We point out that the proposed test for equilibrium compatibility is especially relevant in the recent
debate on the possibility to deduce the non-equilibrium character of a
system from partial observation~\cite{seifert2019stochastic}, in
particular recalling that the time-series of a scalar Gaussian process
(in our case $\omega(t)$) is always symmetric under
time-reversal~\cite{weiss1975time,lucente2022inference}.

\section{Conclusions}
TURs represent an impressive result with
manifold applications, from the evaluation of the entropic cost for
the precision ratio of currents~\cite{pietzonka2017finite}, to the
estimation of entropy production~\cite{manikandan2020inferring} in
non-equilibrium systems, to the identification of limits on the
temporal regimes of anomalous diffusion~\cite{Hartich2021}. 
Considering a class of generalized Langevin equations with several exponential timescales and uniform external force, we have derived a bound for the MSD (Eq.~\eqref{eq::finalBound}) which improves the one obtained through the standard TUR (Eq.~\eqref{TUR_Over}). Indeed, our bound is tighter, valid at all times and useful also for freely diffusing particles. The class of linear models we considered can describe the coupling between relevant degrees of freedom in many-body interacting systems. This allowed us to
test our results on experimental and numerical
data of a tracer diffusing in a granular medium. Moreover, we showed how to use this bound as an immediate tool for inferring non-equilibrium properties of the dynamics from the shape of the MSD. Our approach can be extended to other non-equilibrium systems where several sources of dissipation are present, such as fluids of active particles or driven mixtures. We also recall that linearly coupled equations are the natural framework of linear irreversible thermodynamics, valid (at
small perturbations) also for periodically forced
systems~\cite{brandner2015thermodynamics}. The generalization of our results to non-linear cases such as particles subjected to periodic potentials or non-linear frictional forces represents a promising perspective.  

\acknowledgments{The authors acknowledge the financial support
from the MIUR PRIN2017 project 201798CZLJ. A. Plati acknowledges the financial support by Labex Palm (project FT2AC). The Authors wish to thank Hyunggyu Park for interesting discussions.}

\onecolumngrid
\appendix

\section{Details of calculations for the general model}\label{sec:details}
In this section we report the calculations necessary to obtain some relevant quantities that are used in the main text. For clarity reason we rewrite here the definition of the general model. We consider the multivariate linear stochastic differential equation (SDE) $\dot{\boldsymbol{X}}=\hat{A}\boldsymbol{X}+\hat{B}\boldsymbol{\xi} +\boldsymbol{F}$, where $\boldsymbol{X}=\{\omega,\Omega_1,\ldots,\Omega_n\}$,
$\boldsymbol{\xi}=\{\xi_0,\xi_1,\ldots,\xi_n\}$ and
$\boldsymbol{F}=\{F_{\text{ext}},0,\ldots,0\}$. The interaction and the noise matrices are give by:
\begin{equation}\label{eq::GeneralMatrixApp}
\hat{A}=\left( {\begin{array}{cccc}
    -1/\tau & 1/b_1 & \ldots & 1/b_n \\
   -a_1 b_1 & -1/\tau_1 &0 & 0\\
   \vdots& 0 & \ddots&0 \\
    -a_n b_n& 0 & 0 &-1/\tau_n \\ 
  \end{array} } \right), 
\end{equation}
\begin{equation}\label{eq::GeneralMatrixAppB}
\quad \hat{B}=\text{diag}\left(\sqrt{2q/\tau},\sqrt{2q_1a_1b_1^2/\tau_1},\ldots,\sqrt{2q_na_nb_n^2/\tau_n} \right).
\end{equation} All the model parameters are assumed to be positive. As in the main text, we define $\Delta
\boldsymbol{X}=\boldsymbol{X}-\langle \boldsymbol{X} \rangle$, $\hat{\beta}=\hat{\sigma}^{-1}$ and $\sigma_{ij}=\langle \Delta X_i \Delta X_j \rangle$.

 \subsection{Stationary probability distribution function}\label{sec:statics}

The stationary probability distribution function of the model is the multivariate Gaussian \cite{Gardiner} $P(\boldsymbol{X})\propto \exp\left(- \frac{1}{2} \Delta \boldsymbol{X}
\hat{\beta} \Delta\boldsymbol{X} \right)$. To have an explicit expression of that, one has to solve the following equation for the covariance matrix $\hat\sigma$:
\begin{equation}\label{eq::EqForSigma}
 \hat A\hat\sigma+\hat\sigma  \hat{A}^T=-\hat B \hat{B}^T.
 \end{equation} 
 The solution of such a matrix equation for our model in the general case is cumbersome. Here we report the explicit solution for $n=1$:
 \begin{equation}
  \hat{\sigma}=\frac{1}{(1+a_1\tau\tau_1)(\tau+\tau_1)}\left( {\begin{array}{cc}
    a_1q_1\tau^2\tau_1+q(\tau+\tau_1+a_1\tau\tau_1^2) & a_1b_1 \tau\tau_1 (q_1-q)\\
  a_1b_1\tau\tau_1(q_1-q) & a_1b_1(a_1q\tau\tau_1^2+q_1(\tau+\tau_1+a_1\tau^2\tau_1))  \\
  \end{array} } \right).
  \end{equation}   
Our model is built in such a way to have thermodynamic equilibrium if $q_i=q$ $\forall i$ and $F_{\text{ext}}=0$. In such a condition we expect the equilibrium probability distribution function to be canonical (i.e. $\hat{\beta}^{\text{eq}}_{ij} \propto \delta_{ij}/q$). Now we check that by Eq.~\eqref{eq::EqForSigma}. We assume $\hat{\sigma}^{\text{eq}}_{ij}=c_i\delta_{ij}$ and substitute it into Eq.~\eqref{eq::EqForSigma} with $q_i=q$:
\begin{equation}\label{eq::EqForSigmaConto}
 \left( \hat A\hat\sigma^{\text{eq}}+\hat\sigma^{\text{eq}}  \hat{A}^T\right)_{ij}=A_{ij}c_j+c_iA^T_{ij}=-(B_{ii}^{\text{eq}})^2\delta_{ij}.
\end{equation}  
For $i=j=0$ we have $c_0=-B^2_{00}/A_{00}=2q$ while if $i=j\neq 0$ one has $c_i=-(B_{ii}^{\text{eq}})^2/A_{ii}=2qa_ib_i^2$. With this solutions, is easy to verify that the left hand side of Eq.~\eqref{eq::EqForSigmaConto} is always zero if $i\neq j$. The equilibrium probability distribution function is then given by:

\begin{equation} \label{eq:PDFeq} 
P_{\text{eq}}(\boldsymbol{X})\propto \exp\left[-\frac{1}{2q}\left(\omega^2+\sum_{i=1}^n \frac{\Omega_i^2}{a_ib_i^2}\right)\right].
\end{equation}

\subsection{Entropy production}\label{Sec::EntProd}

We consider the entropy production of the general model defined according to
the Lebowitz and Spohn functional.  We use the relation reported in
\cite{Puglisi2009} that expresses the
entropy production as the product of reversible and irreversible
components of the drift in the Langevin equation. Since we are interested in the entropy production in the stationary state, we only consider the term that is extensive in time.  We obtain 
\begin{eqnarray}
  \Delta S(t)&=& \log\frac{Prob(\{\omega(s),\Omega_1(s),\cdots\Omega_n(s)\}_0^\tau)}{Prob(\{-\omega(t-s),\Omega_1(t-s),\cdots,\Omega_n(t-s)\}_0^t)} \\
  &=&
   \frac{1}{D_\omega}\int_0^t ds~[A^{irr}_\omega(\dot{\omega}(s)-A^{rev}_\omega)]+ \sum_{i=1}^{n}
   \frac{1}{D_{\Omega_i}}\int_0^t ds~[A^{irr}_{\Omega_i}(\dot{\Omega}_i(s)-A^{rev}_{\Omega_i})]
\end{eqnarray}
where $D_\omega=q/\tau, D_{\Omega_i}=q_ia_ib_i^2/\tau_i$ and
\begin{equation}\label{eq::Arevirr}
  A^{rev}_\omega=\sum_{i=1}^n\frac{\Omega_i(s)}{b_i} +F_{\text{ext}}, \quad A^{irr}_\omega=-\frac{\omega(s)}{\tau}, \quad A^{rev}_{\Omega_i}=-a_ib_i\omega(s),\quad A^{irr}_{\Omega_1}=-\frac{\Omega_i(s)}{\tau_i} 
\end{equation}
having used the fact that $\omega$ is odd and the $\Omega_i$s are even under time reversal (see \ref{subs:SymTimeAux} below).
Therefore, for the entropy
production in the stationary state, we obtain
\begin{eqnarray}\label{eq:FluctEntrProd}
\Delta S(t)
&=& \frac{1}{D_\omega} \int_0^t ds~\left(-\frac{\omega(s)}{\tau}\right)\left[\dot{\omega}(s)-\sum_{i=1}^n\frac{\Omega_i(s)}{b_i}-F_{\text{ext}}\right]+
\sum_{i=1}^{n}
   \frac{1}{D_{\Omega_i}}\int_0^t ds~\left(-\frac{\Omega_i(s)}{\tau_i}\right)\left[\dot{\Omega}_i(s)+a_ib_i\omega(s)\right] \\
&=& \frac{1}{D_\omega}\left[-\frac{\delta \omega^2}{2\tau} +\frac{1}{\tau}\int_0^t ds~\omega(s)\left(\sum_{i=1}^n\frac{\Omega_i(s)}{b_i}+F_{\text{ext}}\right)\right]+
\sum_{i=1}^n\frac{1}{D_{\Omega_i}}\left[-\frac{\delta \Omega_i^2}{2\tau_i} -\frac{a_ib_i}{\tau_i}\int_0^t ds~\omega(s)\Omega_i(s)\right] \\ \label{eq:FluctEntrProdC}
&=& -\frac{\delta_t (\omega^2)}{2q}  -\sum_{i=1}^n \frac{\delta_t (\Omega_i^2)}{2q_ia_ib_i^2} + \frac{1}{q}\int_0^t ds~\omega(s) F_{\text{ext}} + \sum_{i=1}^n \frac{1}{b_i}\int_0^t ds~\left(\frac{1}{q}-\frac{1}{q_1}\right)\omega(s)\Omega_i(s)
\end{eqnarray}
where we introduced the notation $\delta_t (z)=z(t)-z(0)$. Considering that in the stationary state we expect $\langle \Delta S(t) \rangle=\langle \dot{S}\rangle t$, the average entropy production rate is then
\begin{equation}  \label{eq:EntProdRate}
\langle \dot{S}\rangle = \frac{1}{q}\langle \omega\rangle F_{\text{ext}} + \sum_{i=1}^n \frac{1}{b_i}\left(\frac{1}{q}-\frac{1}{q_i}\right)\langle\omega\rangle\langle\Omega_i\rangle + \sum_{i=1}^n \frac{1}{b_i}\left(\frac{1}{q}-\frac{1}{q_i}\right)\sigma_{0i}
\end{equation}
that coincides with the expression reported in Eq.~\eqref{eq::EntProdTOTALgen} of the main text. 
It is also important to note that Eq.~\eqref{eq:FluctEntrProdC} is consistent with thermodynamic interpretation for which, at equilibrium, the only contribute to the fluctuating entropy production is the work done by the thermal bath. Indeed, rescaling the auxiliary variables as $\tilde{\Omega}_i=\Omega_i/\sqrt{a_ib_i^2}$, one obtains:\begin{equation} \label{eq:checkEntProdEq}
 \Delta S^{\text{eq}}(t)=-\frac{\delta_t (\omega^2)}{2q}  -\sum_{i=1}^n \frac{\delta_t (\tilde{\Omega}_i^2)}{2q} 
\end{equation}
that is evidently zero when averaged on the stationary state. The interpretation of the above expression as the total fluctuating work done by the thermostats is consistent with the equilibrium probability distribution function (Eq.~\eqref{eq:PDFeq}).

\subsection{Equilibrium condition for the Fokker-Planck equation}\label{subs:eqFPE}
The Fokker-Planck equation associated to Eq.~\eqref{model} reads:
\begin{equation}
\partial_t P(\boldsymbol{X},t)=-\nabla\cdot\left(\boldsymbol{J}^{\text{rev}}(\boldsymbol{X},t)+\boldsymbol{J}^{\text{irr}}(\boldsymbol{X},t)\right)
\end{equation} where:
\begin{equation}
J_i^{\text{irr}}(\boldsymbol{X},t)=\left[A_i^{\text{irr}}(\boldsymbol{X},t)-\frac{1}{2}B^2_{ii}\partial_{X_i}\right]P(\boldsymbol{X},t)
\end{equation}
\begin{equation}
J_i^{\text{rev}}(\boldsymbol{X},t)=A_i^{\text{rev}}(\boldsymbol{X},t)P(\boldsymbol{X},t).
\end{equation}
From Eq.~\eqref{eq::GeneralMatrixAppB} and \eqref{eq::Arevirr} 
it is easy to check that, as we expect, the probability distribution function given by Eq.~\eqref{eq:PDFeq} makes the irreversibe current $J_i^{\text{irr}}(\boldsymbol{X},t)$ equal to zero when $q_i=q$ $\forall i$.

\subsection{Symmetry under time reversal of the auxiliary variables}\label{subs:SymTimeAux}
\begin{figure}
\includegraphics[width=0.59\textwidth]{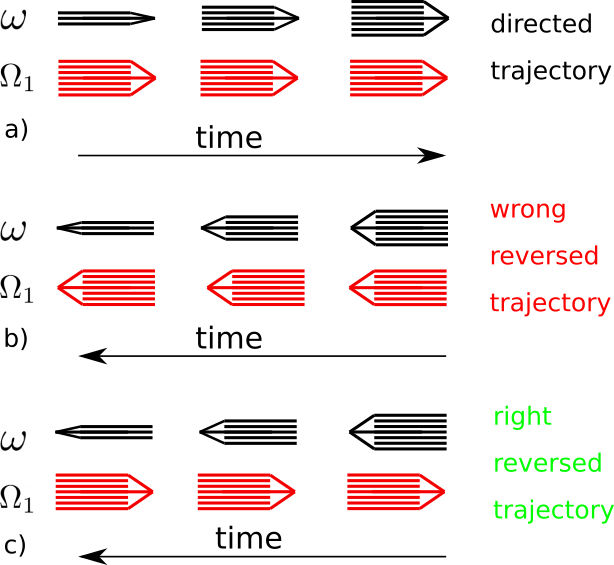}\caption{Evolution of $\omega$ and $\Omega_1$, the arrow width corresponds to vector magnitude. a) directed trajectories of $\omega$ and $\Omega_1$: the auxiliary variable increases the intruder's velocity. We consider a limit in which $\Omega_1$ is not perturbed by the intruder to ease the readability of the cartoon. b) Reversed trajectories with both the variables considered odd under time reversal. Here the auxiliary field would naturally increase the intruder's velocity so the observed slowing down is entirely originated by the action of noise. c) Reversed trajectories considering $\omega$ odd and $\Omega_1$ even under time reversal. The auxiliary variable increases the intruder's velocity with the same probability.}   \label{fig:Comparison_1D}
\end{figure}

The calculations done so far assume the auxiliary variables to be even under time reversal. This is a forced choice if we want to obtain the correct thermodynamic interpretation expressed by Eq. \eqref{eq:checkEntProdEq}. Nevertheless, this choice may seem unphysical because in our model the $\Omega_i$s and $\omega$ can have the same physical dimensions (see Appendix~\ref{sec::Fit} below) so one expects them to follow the same symmetry under time reversal. Here we want to provide an argument that clarifies why considering even $\Omega_i$s is actually reasonable from a physical point of view. Let's consider a particular case of our general model where $n=1$, $q=q_1$, $\alpha=0$, $F_{\text{ext}}=0$ and $\tau_1\gg \tau$. The equations of motion then read:
\begin{equation}
 \dot{\omega}=-\frac{1}{\tau}(\omega-\Omega_1)+\sqrt{\frac{2q}{\tau}}\xi_0, \quad  \dot{\Omega}_1=0 .
\end{equation} 
These equations represent the diffusion of an intruder ($\omega$) in a fluid with a local velocity field ($\Omega_1$) that relaxes on timescales much larger than $\tau$. If the two variables have the same(opposite) sign the velocity field fastens(slows down) the intruder.  Being a sub-case of the general model with $q=q_1$ (i.e. thermodynamic equilibrium), we expect for the trajectories of $\{\omega(t),\Omega_1(t)\}$ to have the same probability under time reversal. In Fig.~\ref{fig:Comparison_1D}, we show one possible directed evolution of the two variables and the comparison between time reversal operations where $\Omega_1$ is considered odd or even. It is clear that the case in which $\Omega_1$ is odd (b) requires a (very improbable) realization of the noise that is able to slow down $\omega$ despite the positive contribute of $\Omega_1$. On the contrary, it is reasonable to think that the reversed trajectories with even $\Omega_1$ (c) can be obtained with a realization of the noise that has the same probability of the directed one. The conclusion we draw from this cartoon is that we must consider auxiliary variables as external fields. Thus, we don't change their sign under time reversal even if they have the same physical dimension of a velocity and they are influenced by the intruder dynamics.

\subsection{Diffusion coefficient}\label{sec::diffcoeff}
To obtain general expression of the diffusion coefficient of our model: $\lim_{t\to \infty}\langle \Delta\theta(t)^2 \rangle/t=\mathcal{S}_{00}(0)=\left[\hat A^{-1}\hat B\hat B^T(\hat A^T)^{-1}\right]_{00}$ one needs to invert the arrowhead matrix $\hat{A}^T$. We first perform the matrix product and get:
\begin{equation}
\left[\hat A^{-1}\hat B \hat B^T(\hat A^T)^{-1}\right]_{00}=\sum_{kj}A^{-1}_{0j}B^2_{jj}\delta_{jk}A^{-1}_{0k}=\sum_k \left(A^{-1}_{0k}B_{kk}\right)^2,
\end{equation}
where the sums run from $0$ to $n+1$ and $\delta_{jk}$ is the Kronecker delta. From Ref. \citep{Wani2016} we know that:
\begin{equation}
\det(A)A^{-1}_{00}=(-1)^n\prod_{j=1}^n \frac{1}{\tau_j}, \quad \det(A)A^{-1}_{0k}=(-1)^n\frac{\tau_k}{b_k}\prod_{j=1}^n \frac{1}{\tau_j}, \quad \det(A)=(-1)^{n+1}\prod_{j=1}^n \frac{1}{\tau_j}\left[ \frac{1}{\tau} +\sum_{i=1}^na_i\tau_i \right]
\end{equation}
and with some algebraic manipulation we arrive to 
\begin{equation}\label{eq:generalrDiff}
\mathcal{S}_{00}(0)=\mathcal{D}_{\text{eq}}\left[\frac{1+\tau\sum_k \frac{q_k}{q} a_k\tau_k}{1+\tau\sum_k a_k\tau_k}\right]
\end{equation}
that coincides with Eq.~\eqref{eq::generalDiffCoeff} of the main text. 

\subsection{TUR in the large time limit}\label{sec::turlargetimeapp}
The TUR valid at large times reported in Eq.~\eqref{eq:TurOverBetter} 
of the  main text has been derived by directly evaluating the quantities involved in it from the general model: 
\begin{equation}\label{eq:TurOverBetterSM}
\lim_{t\to \infty}\frac{\langle \Delta\theta(t)^2 \rangle}{t}\ge \frac{2\langle \omega \rangle^2}{\langle \dot{S}\rangle_{\text{ext}}}\ge \frac{2\langle \omega \rangle^2}{\langle \dot{S}\rangle}.
\end{equation}
The first inequality follows from verifying that:
\begin{equation}
\left[ \frac{1+\tau\sum_k \frac{q_k}{q}a_k\tau_k}{1+\tau\sum_k a_k \tau_k} \right]-\left[\frac{1+\tau\sum_k a_k \tau_k}{1+\tau\sum_k \frac{q}{q_k}a_k\tau_k} \right]=\tau\sum_k\frac{a_k\tau_k}{qq_k}   \left(q-q_k\right)^2+\tau^2\sum_{j>k}\frac{\tau_ja_j\tau_ka_k}{q_kq_j}\left(q_j-q_k\right)^2\ge 0.
\end{equation}
The second inequality of \eqref{eq:TurOverBetterSM} is directly related to the decomposition of the entropy production rate $\langle \dot{S}\rangle=\langle \dot{S}\rangle_{\textrm{ext}}+\langle \dot{S}\rangle_{\textrm{th}}$ and the positivity of $\langle \dot{S}\rangle_{\textrm{th}}$. We recall that $\langle \dot{S}\rangle_{\textrm{th}}=\sum_i \frac{1}{b_i} (\frac{1}{q} -
\frac{1}{q_i})\hat{\sigma}_{0i} \ge 0$ follows from the fact that in absence of external force it is the only contribute to the entropy production rate and that its expression does not depend on $F_{\text{ext}}$ thanks to the linearity of the model.

\section{Underdamped TUR from Cramér-Rao inequality}\label{secc::cramerraoapp}
\subsection{Relation with previously derived TUR}
The derivation of the TUR in the large time limit has been performed exploiting the fact that it is possible to derive an explicit and compact expression for both the diffusion coefficient (Eq.~\eqref{eq:generalrDiff}) and the entropy production rate (Eq.~\eqref{eq:EntProdRate}) in our model. Using the same procedure to derive a TUR valid at all timescales is much more complicated because: i) one has to handle the general expression of the MSD of the model that is cumbersome, ii) one has to guess a time-dependent functional form of the bound.

A TUR valid at all timescales for a general Langevin dynamics with a fully underdamped structure has been derived in \citep{Lee2021} following the method explained in \citep{Hasegawa2019II}. With fully underdamped we mean a system where one half of the degrees of freedom is even under time reversal and is obtained as the derivative of the other half that is odd under time reversal. Such a TUR takes the following form:
\begin{equation}\label{eq:LeeParkTur}
 \frac{\text{Var}(\Theta(t))}{\langle \Theta(t) \rangle^2} \ge \frac{1}{\Delta S (t) + \mathcal{I}}
\end{equation}
where $\Theta(t)$ is a generalized integrated current, $\Delta S (t)$ is the \emph{total} entropy production and $\mathcal{I}$ is a non-extensive term in time. It is worth mentioning that \eqref{eq:LeeParkTur} represents an improvement with respect the underdamped TUR derived in \cite{Hasegawa2019} because it has the correct large time limit. With some calculations (not shown) it is possible to show that the same TUR can be derived also for our model that has a partial underdamped structure (i.e. $\omega$ is odd and all the $\Omega_i$s are even under time reversal). Nevertheless, the above TUR in the large time limit brings to an inequality that can be improved by substituting $\Delta S (t)$ with $\Delta S_{\text{ext}} (t)$. As reported in the main text, one of the main results of our work is the derivation of the following TUR:
\begin{equation}\label{eq:PPS}
\frac{\langle \Delta\theta(t)^2 \rangle}{(\langle \omega \rangle t)^2} \ge \frac{2}{\Delta S_{\text{ext}}(t)+\mathcal{I}}
\end{equation}
It is valid at all times in the steady state and brings to the improved inequality \eqref{eq:TurOverBetterSM} in the large time limit.

\subsection{Details of the derivation}\label{sec::crdetails}
In order to derive the TUR \eqref{eq:PPS} from the Cramér-Rao inequality (Eq.~\eqref{eq::CramRao} in the main text) we write the SDE of our model with a perturbation depending on the parameter $h$:
\begin{equation}
dX_i=f_i^{h}(\boldsymbol{X})dt+B_{ii}dW(t)
\end{equation}
where $dW(t)$ is the increment of the Wiener process and:
\begin{equation}\label{eq:FictDrift}
f_i^h(\boldsymbol{X})=\sum_j A_{ij} X_j+F_i+h V_i .
\end{equation}
We then apply the main results of Ref. \citep{Hasegawa2019II}. Considering initial conditions in the steady state, the Fisher information takes the following form
\begin{equation}
\mathcal{I_{\text{F}}}(h)=-\langle \partial^2_h \ln P_h(\boldsymbol{X}) \rangle_h + \left\langle \int_0^t dt'\sum_i \left( \frac{\partial_h f_i^{h}(\boldsymbol{X})}{B_{ii}}   \right)^2 \right\rangle_h
\end{equation}
where $P_h(\boldsymbol{X})$ is the probability distribution function of the perturbed process and $\langle \cdot \rangle_h$ refers to averages on such a probability.  Since the system is linear and the $h$-perturbation does not depend on $\boldsymbol{X}$, the stationary probability distribution associated to the fictive dynamics is still a multivariate Gaussian with the same covariance matrix but different average values $\langle X_i \rangle_h$. Thus, $P_h (\boldsymbol{X})\propto \exp \left[ \right(\boldsymbol{X}-\langle \boldsymbol{X} \rangle_h)^T \hat{\beta} (\boldsymbol{X}-\langle \boldsymbol{X} \rangle_h)/2]$. Moreover, from Eq.~\eqref{eq:FictDrift} one has: $\partial_h f_i^{h}(\boldsymbol{X})=V_i$. In order to make the Cramér-Rao inequality fully explicit, we then need to compute the average values $\langle X_i \rangle_h$. With the specific choice done in the main text $\boldsymbol{V}=\{\langle \omega \rangle/\tau,-\langle \Omega_1
\rangle/\tau_1,\ldots,-\langle \Omega_n \rangle/\tau_n\}$, the following relations must be satisfied:
\begin{subequations}
\begin{equation}
-\frac{1}{\tau}\langle \omega\rangle_h + \sum_i \frac{\langle \Omega_i \rangle_h}{b_i}+F_{\text{ext}}+hV_0=0
\end{equation}
\begin{equation}
-\frac{1}{\tau_i}\langle \Omega_i \rangle_h-a_ib_i\langle \omega\rangle_h-hV_i=0
\end{equation}
\end{subequations}
from which we obtain $\langle \omega \rangle_h=(1+h)\langle \omega \rangle$ and $\langle \Omega_i \rangle_h=\langle \Omega_i \rangle$.

Substituting these relations in the Cramér-Rao inequality for $h=0$ we find the TUR \eqref{eq:PPS}. Indeed the Fisher information becomes
\begin{equation}\label{eq:FisherAlmostDone}
\mathcal{I_{\text{F}}}(0)=\int d\boldsymbol{X} \frac{[\partial_h P_h(\boldsymbol{X})]^2|_{h=0}}{P( \boldsymbol{X})}+ \frac{1}{2}\left[ \frac {\langle \omega \rangle^2}{\tau q} + \sum_i \frac{\langle \Omega_i \rangle^2}{\tau _i q_i a_i b_i^2} \right] t = \frac{1}{2}\left[\mathcal{I}+ \langle \dot{S} \rangle_{\text{ext}}t \right].
\end{equation}
The relation between the first term of Eq.~\eqref{eq:FisherAlmostDone} and Eq.~\eqref{eq::nonExt} of the main text follows from the direct evaluation of the probability distribution's derivative with respect $h$:
\begin{equation}
\int d\boldsymbol{X} \frac{[\partial_h P_h(\boldsymbol{X})]^2|_{h=0}}{P( \boldsymbol{X})}=\left\langle \left[\frac{1}{2}\sum_{jn}\beta_{jn}\partial_h(\Delta X_j^{h}\Delta X_n^{h})\right]^2\right\rangle=\langle \omega \rangle^2\sum_{jn}\beta_{0n}\beta_{0j}\sigma_{nj}=\langle \omega \rangle^2\beta_{00},
\end{equation}
where we used $\Delta X_j^{h}=(X_j-\langle X_j \rangle_h)$ and $\partial_h \Delta X_j^{h}=\langle\omega \rangle\delta_{0j}$.
Finally, using the relation between $\langle\omega \rangle$ and $\langle\Omega_i\rangle$ (Eq.~\eqref{mean} of the main text), we note that:
\begin{equation}
\frac{\langle\omega \rangle^2}{\tau q}=\frac{1}{q}\omega F_{\text{ext}}+\sum_i \frac{\langle\omega\rangle \langle\Omega_i\rangle}{q b_i} \quad \text{and} \quad  \frac{\langle \Omega_i \rangle^2}{\tau _i q_i a_i b_i^2}=-\frac{\langle \omega\rangle\langle \Omega_i \rangle}{q_ib_i}.
\end{equation}
So, we find that the second term of Eq.~\eqref{eq:FisherAlmostDone} is directly related to the entropy production rate as expressed in Eq.~\eqref{eq::EntProdTOTALgen}.

\section{Fitting procedure} \label{sec::Fit}
 
In order to fit the model's parameter we used two distinct methods for numerical and experimental data. In the numerical data, independent measurements of the auto-correlation and the response function of the granular intruder are available \cite{sarracino2010irreversible}. 
The model we used for them is defined by the following matrices:
\begin{equation}\label{eq::Mat2d}
\hat{A}=\left( {\begin{array}{cc}
    -1/\tau & 1/\tau \\
   -\alpha/\tau_1 & -1/\tau_1  \\
  \end{array} } \right), \quad \hat{B}=\left( {\begin{array}{cc}
    \sqrt{2q/\tau} & 0 \\
   0 & \sqrt{2q_1\alpha\tau/\tau_1^2}  \\
  \end{array} } \right).
\end{equation}
so it counts five parameters $\tau$, $q$, $\alpha$, $\tau_1$, $q_1$. A multi-branch fit of auto-correlation and response allows to determine the numerical value of such parameters without overfitting. Regarding the experiments performed at moderate density (reported in Fig.~\ref{fig:Fig1}a of the main text) we still use \eqref{eq::Mat2d} but here we have only data with which we can reconstruct the autocorrelation, the MSD and the power spectral density of the velocity (PSDV) in the steady state. These are all observables that store the same amount of information in different ways. Indeed, knowing the autocorrelation function, we can obtain the MSD with the Kubo's formula or the PSDV by a Fourier transform.
In a linear model with $n+1$ variables, the autocorrelation function is a sum of $n+1$ exponential decays each one identified by an amplitude and a characteristic time. Thus, a fit of the autocorrelation or an equivalent observable alone, can be used to estimate a maximum of $2(n+1)$ parameters. In order to have four free parameters, we have fixed $\alpha=1$ before doing the fit of the experimental data at moderate density. A similar procedure has to be done to fit the experimental data at high density (shown in Fig.~\ref{fig:Fig1}b of the main  text). In this case the matrices of the model are given by:
\begin{equation}
\hat A=\left( {\begin{array}{ccc}
    -1/\tau & 1/\tau & 1/\tau \\
   -\alpha/\tau_1 & -1/\tau_1  & 0\\
   -\epsilon^2/\tau_2& 0 & -1/\tau_2 \\
  \end{array} } \right), \quad \hat B=\left( {\begin{array}{ccc}
    \sqrt{2q/\tau} & 0 & 0\\
   0 & \sqrt{2q_1\alpha\tau/\tau_1^2}  & 0\\
   0& 0 & \epsilon^{3/2}\sqrt{2q_2/\tau_2} \\
  \end{array} } \right).
\end{equation}
Remembering that $\epsilon=\tau/\tau_2$, we have seven parameters $\tau$, $q$, $\alpha$, $\tau_1$, $q_1$, $\tau_2$, $q_2$  and in order to not overfit we fixed $q=q_1$ before doing the fit. 

With this fitting procedure we are able to reproduce the MSD and the PSDV (not shown) but dealing with a large number of parameters we know that there is probably an entire region of the parameter space where we could find a good agreement with the experimental data. In view of this, we stress that the important point of our analysis  is that there is a set of parameters well reproducing our data for which is important to take into account the correct terms of EPR in the TUR. 
Nevertheless, it is also important to note that the arbitrariness in the estimate of model's parameters from data is a quite general issue. In light of this, we remark that the last result presented in the main text (i.e. non-equilibrium signatures in the shape of the MSD) does not require any fit of the data.

\section{Extent of the anomalous diffusion}\label{sec::boundonext}
Here we want to adapt the analysis done in \cite{Hartich2021} to the new bound derived in the main text. Considering a regime where the MSD behave as $\langle \Delta\theta(t)^2 \rangle\sim K_{\nu}t^{\nu}$ we have that: 
\begin{equation}
K_{\nu}t^{\nu} \ge \frac{C_1 t^2}{1+C_2 t} \sim \begin{cases}
C_1t^2 \quad t\to 0\\
\frac{C_1}{C_2} t \quad t\to \infty
\end{cases}
\end{equation}
where $C_1=2\langle \omega \rangle^2/\mathcal{I}$ and $C_2=\langle \dot{S}\rangle_{\text{ext}}/\mathcal{I}$. The above inequality is satisfied only for times that solve $t^{\nu-2}+C_2t^{\nu-1}-C_1/K_\nu \ge 0$. We can take $\nu=0$ for an example of the subdiffusive case and $\nu=2$ for the superdiffusive one obtaining:

\begin{equation}
t^*_{\text{sub}} \le \frac{2}{C_2}\left( \sqrt{1+\frac{4C_1}{C_2^2K_0}} \right)^{-1}, \quad t^*_{\text{super}} \ge \frac{1}{C_2}\left( \frac{C_1}{K_2}-1 \right).
\end{equation}
We note that in the subdffusive case there is always a positive time that prevents the extension of the subdiffusion after a certain time. On the other hand, the superdiffusive one has a meaningful bound only if $K_2 < C_1$ i.e. if the anomalous diffusion coefficient is lower than the ballistic one of the bound. Having in mind a loglog plot, it means that if the superdiffusive regime $K_2 t^2$ lays over the line $C_1 t^2$ it can holds for any positive times. In the opposite case the onset of such regime can not occur before $t^*_{\text{super}}$. This is consistent with the fact that the ballistic regime is always present in an underdamped system for $t \sim 0$. The bound applies to the anomalous superdiffusive regimes that may appear at larger times as the one shown in Fig.~\ref{fig:Fig1}b of the main text.

\begin{figure}[h]
\includegraphics[width=0.5\textwidth]{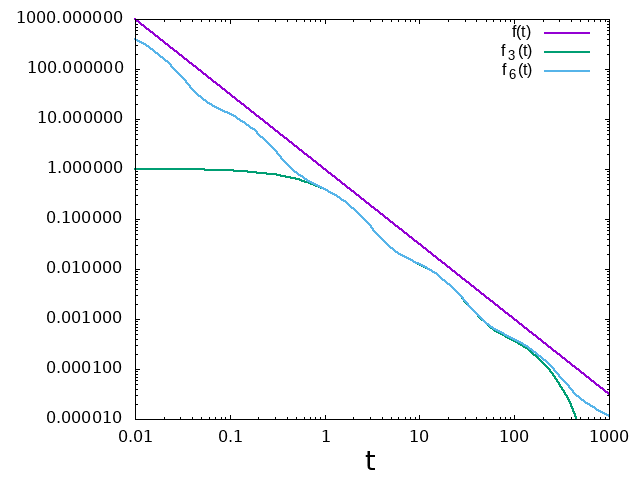}
\caption{Comparison between a power law decay $f(t)$ and two approximations given by sum of exponentials. See text for the definition of the parameters. \label{fig:powerlaw}}
\end{figure}

\section{Power law decay and sum of exponentials}\label{sec:powerLaw}

It is interesting to realize that the choice of a memory kernel which is sum of exponentials with different decay rates can reproduce physical situations where memory decays as a power law, of course with a maximum time cut-off. We are not able to provide a general theory, but visual examples constitute an empirical proof. In Fig.~\ref{fig:powerlaw} we compare the following three decaying functions of time $t$:
\begin{align}
f(x)=t^{-3/2}\\
f_3(x)=\sum_{k=1}^3 \frac{a_k}{\tau_k}e^{-t/\tau_k}\\
f_6(x)=\sum_{k=1}^6 \frac{a_k}{\tau_k}e^{-t/\tau_k}\\
\textrm{with} \;\;\; a_k =\tau_k^{-1/2} \;\;\; \textrm{and} \;\;\; \tau \equiv\{1,10,100,0.01,0.1,1000\}.
\end{align}
A more systematic study about how to use exponential functions to approximate power laws is also provided in \cite{Bochud2007}.

\twocolumngrid

\bibliographystyle{apsrev4-1}
\bibliography{biblioTurnew}

\begin{thebibliography}{47}%
\makeatletter
\providecommand \@ifxundefined [1]{%
 \@ifx{#1\undefined}
}%
\providecommand \@ifnum [1]{%
 \ifnum #1\expandafter \@firstoftwo
 \else \expandafter \@secondoftwo
 \fi
}%
\providecommand \@ifx [1]{%
 \ifx #1\expandafter \@firstoftwo
 \else \expandafter \@secondoftwo
 \fi
}%
\providecommand \natexlab [1]{#1}%
\providecommand \enquote  [1]{``#1''}%
\providecommand \bibnamefont  [1]{#1}%
\providecommand \bibfnamefont [1]{#1}%
\providecommand \citenamefont [1]{#1}%
\providecommand \href@noop [0]{\@secondoftwo}%
\providecommand \href [0]{\begingroup \@sanitize@url \@href}%
\providecommand \@href[1]{\@@startlink{#1}\@@href}%
\providecommand \@@href[1]{\endgroup#1\@@endlink}%
\providecommand \@sanitize@url [0]{\catcode `\\12\catcode `\$12\catcode
  `\&12\catcode `\#12\catcode `\^12\catcode `\_12\catcode `\%12\relax}%
\providecommand \@@startlink[1]{}%
\providecommand \@@endlink[0]{}%
\providecommand \url  [0]{\begingroup\@sanitize@url \@url }%
\providecommand \@url [1]{\endgroup\@href {#1}{\urlprefix }}%
\providecommand \urlprefix  [0]{URL }%
\providecommand \Eprint [0]{\href }%
\providecommand \doibase [0]{http://dx.doi.org/}%
\providecommand \selectlanguage [0]{\@gobble}%
\providecommand \bibinfo  [0]{\@secondoftwo}%
\providecommand \bibfield  [0]{\@secondoftwo}%
\providecommand \translation [1]{[#1]}%
\providecommand \BibitemOpen [0]{}%
\providecommand \bibitemStop [0]{}%
\providecommand \bibitemNoStop [0]{.\EOS\space}%
\providecommand \EOS [0]{\spacefactor3000\relax}%
\providecommand \BibitemShut  [1]{\csname bibitem#1\endcsname}%
\let\auto@bib@innerbib\@empty
\bibitem [{\citenamefont {Cavagna}(2009)}]{cavagna2009supercooled}%
  \BibitemOpen
  \bibfield  {author} {\bibinfo {author} {\bibfnamefont {A.}~\bibnamefont
  {Cavagna}},\ }\href@noop {} {\bibfield  {journal} {\bibinfo  {journal}
  {Physics Reports}\ }\textbf {\bibinfo {volume} {476}},\ \bibinfo {pages} {51}
  (\bibinfo {year} {2009})}\BibitemShut {NoStop}%
\bibitem [{\citenamefont {Marty}\ and\ \citenamefont
  {Dauchot}(2005)}]{marty2005subdiffusion}%
  \BibitemOpen
  \bibfield  {author} {\bibinfo {author} {\bibfnamefont {G.}~\bibnamefont
  {Marty}}\ and\ \bibinfo {author} {\bibfnamefont {O.}~\bibnamefont
  {Dauchot}},\ }\href@noop {} {\bibfield  {journal} {\bibinfo  {journal}
  {Physical review letters}\ }\textbf {\bibinfo {volume} {94}},\ \bibinfo
  {pages} {015701} (\bibinfo {year} {2005})}\BibitemShut {NoStop}%
\bibitem [{\citenamefont {Bodrova}\ \emph {et~al.}(2012)\citenamefont
  {Bodrova}, \citenamefont {Dubey}, \citenamefont {Puri},\ and\ \citenamefont
  {Brilliantov}}]{bodrova2012intermediate}%
  \BibitemOpen
  \bibfield  {author} {\bibinfo {author} {\bibfnamefont {A.}~\bibnamefont
  {Bodrova}}, \bibinfo {author} {\bibfnamefont {A.~K.}\ \bibnamefont {Dubey}},
  \bibinfo {author} {\bibfnamefont {S.}~\bibnamefont {Puri}}, \ and\ \bibinfo
  {author} {\bibfnamefont {N.}~\bibnamefont {Brilliantov}},\ }\href@noop {}
  {\bibfield  {journal} {\bibinfo  {journal} {Physical Review Letters}\
  }\textbf {\bibinfo {volume} {109}},\ \bibinfo {pages} {178001} (\bibinfo
  {year} {2012})}\BibitemShut {NoStop}%
\bibitem [{\citenamefont {Scalliet}\ \emph {et~al.}(2015)\citenamefont
  {Scalliet}, \citenamefont {Gnoli}, \citenamefont {Puglisi},\ and\
  \citenamefont {Vulpiani}}]{scalliet2015cages}%
  \BibitemOpen
  \bibfield  {author} {\bibinfo {author} {\bibfnamefont {C.}~\bibnamefont
  {Scalliet}}, \bibinfo {author} {\bibfnamefont {A.}~\bibnamefont {Gnoli}},
  \bibinfo {author} {\bibfnamefont {A.}~\bibnamefont {Puglisi}}, \ and\
  \bibinfo {author} {\bibfnamefont {A.}~\bibnamefont {Vulpiani}},\ }\href@noop
  {} {\bibfield  {journal} {\bibinfo  {journal} {Physical review letters}\
  }\textbf {\bibinfo {volume} {114}},\ \bibinfo {pages} {198001} (\bibinfo
  {year} {2015})}\BibitemShut {NoStop}%
\bibitem [{\citenamefont {Plati}\ and\ \citenamefont
  {Puglisi}(2020)}]{plati2020slow}%
  \BibitemOpen
  \bibfield  {author} {\bibinfo {author} {\bibfnamefont {A.}~\bibnamefont
  {Plati}}\ and\ \bibinfo {author} {\bibfnamefont {A.}~\bibnamefont
  {Puglisi}},\ }\href@noop {} {\bibfield  {journal} {\bibinfo  {journal}
  {Physical Review E}\ }\textbf {\bibinfo {volume} {102}},\ \bibinfo {pages}
  {012908} (\bibinfo {year} {2020})}\BibitemShut {NoStop}%
\bibitem [{\citenamefont {Plati}\ \emph {et~al.}(2019)\citenamefont {Plati},
  \citenamefont {Baldassarri}, \citenamefont {Gnoli}, \citenamefont
  {Gradenigo},\ and\ \citenamefont {Puglisi}}]{plati2019dynamical}%
  \BibitemOpen
  \bibfield  {author} {\bibinfo {author} {\bibfnamefont {A.}~\bibnamefont
  {Plati}}, \bibinfo {author} {\bibfnamefont {A.}~\bibnamefont {Baldassarri}},
  \bibinfo {author} {\bibfnamefont {A.}~\bibnamefont {Gnoli}}, \bibinfo
  {author} {\bibfnamefont {G.}~\bibnamefont {Gradenigo}}, \ and\ \bibinfo
  {author} {\bibfnamefont {A.}~\bibnamefont {Puglisi}},\ }\href@noop {}
  {\bibfield  {journal} {\bibinfo  {journal} {Physical review letters}\
  }\textbf {\bibinfo {volume} {123}},\ \bibinfo {pages} {038002} (\bibinfo
  {year} {2019})}\BibitemShut {NoStop}%
\bibitem [{\citenamefont {Lasanta}\ and\ \citenamefont
  {Puglisi}(2015)}]{lasanta2015itinerant}%
  \BibitemOpen
  \bibfield  {author} {\bibinfo {author} {\bibfnamefont {A.}~\bibnamefont
  {Lasanta}}\ and\ \bibinfo {author} {\bibfnamefont {A.}~\bibnamefont
  {Puglisi}},\ }\href@noop {} {\bibfield  {journal} {\bibinfo  {journal} {The
  Journal of Chemical Physics}\ }\textbf {\bibinfo {volume} {143}},\ \bibinfo
  {pages} {064511} (\bibinfo {year} {2015})}\BibitemShut {NoStop}%
\bibitem [{\citenamefont {Seifert}(2019)}]{seifert2019stochastic}%
  \BibitemOpen
  \bibfield  {author} {\bibinfo {author} {\bibfnamefont {U.}~\bibnamefont
  {Seifert}},\ }\href@noop {} {\bibfield  {journal} {\bibinfo  {journal}
  {Annual Review of Condensed Matter Physics}\ }\textbf {\bibinfo {volume}
  {10}},\ \bibinfo {pages} {171} (\bibinfo {year} {2019})}\BibitemShut
  {NoStop}%
\bibitem [{\citenamefont {Barato}\ and\ \citenamefont
  {Seifert}(2015)}]{Barato2015}%
  \BibitemOpen
  \bibfield  {author} {\bibinfo {author} {\bibfnamefont {A.~C.}\ \bibnamefont
  {Barato}}\ and\ \bibinfo {author} {\bibfnamefont {U.}~\bibnamefont
  {Seifert}},\ }\href {\doibase 10.1103/PhysRevLett.114.158101} {\bibfield
  {journal} {\bibinfo  {journal} {Phys. Rev. Lett.}\ }\textbf {\bibinfo
  {volume} {114}},\ \bibinfo {pages} {158101} (\bibinfo {year}
  {2015})}\BibitemShut {NoStop}%
\bibitem [{\citenamefont {Seifert}(2018)}]{seifert2018stochastic}%
  \BibitemOpen
  \bibfield  {author} {\bibinfo {author} {\bibfnamefont {U.}~\bibnamefont
  {Seifert}},\ }\href@noop {} {\bibfield  {journal} {\bibinfo  {journal}
  {Physica A: Statistical Mechanics and its Applications}\ }\textbf {\bibinfo
  {volume} {504}},\ \bibinfo {pages} {176} (\bibinfo {year}
  {2018})}\BibitemShut {NoStop}%
\bibitem [{\citenamefont {Gingrich}\ \emph {et~al.}(2016)\citenamefont
  {Gingrich}, \citenamefont {Horowitz}, \citenamefont {Perunov},\ and\
  \citenamefont {England}}]{Gingrich2016}%
  \BibitemOpen
  \bibfield  {author} {\bibinfo {author} {\bibfnamefont {T.~R.}\ \bibnamefont
  {Gingrich}}, \bibinfo {author} {\bibfnamefont {J.~M.}\ \bibnamefont
  {Horowitz}}, \bibinfo {author} {\bibfnamefont {N.}~\bibnamefont {Perunov}}, \
  and\ \bibinfo {author} {\bibfnamefont {J.~L.}\ \bibnamefont {England}},\
  }\href {\doibase 10.1103/PhysRevLett.116.120601} {\bibfield  {journal}
  {\bibinfo  {journal} {Phys. Rev. Lett.}\ }\textbf {\bibinfo {volume} {116}},\
  \bibinfo {pages} {120601} (\bibinfo {year} {2016})}\BibitemShut {NoStop}%
\bibitem [{\citenamefont {Van~Vu}\ and\ \citenamefont
  {Hasegawa}(2019)}]{Hasegawa2019}%
  \BibitemOpen
  \bibfield  {author} {\bibinfo {author} {\bibfnamefont {T.}~\bibnamefont
  {Van~Vu}}\ and\ \bibinfo {author} {\bibfnamefont {Y.}~\bibnamefont
  {Hasegawa}},\ }\href {\doibase 10.1103/PhysRevE.100.032130} {\bibfield
  {journal} {\bibinfo  {journal} {Phys. Rev. E}\ }\textbf {\bibinfo {volume}
  {100}},\ \bibinfo {pages} {032130} (\bibinfo {year} {2019})}\BibitemShut
  {NoStop}%
\bibitem [{\citenamefont {Hasegawa}\ and\ \citenamefont
  {Van~Vu}(2019)}]{Hasegawa2019II}%
  \BibitemOpen
  \bibfield  {author} {\bibinfo {author} {\bibfnamefont {Y.}~\bibnamefont
  {Hasegawa}}\ and\ \bibinfo {author} {\bibfnamefont {T.}~\bibnamefont
  {Van~Vu}},\ }\href {\doibase 10.1103/PhysRevE.99.062126} {\bibfield
  {journal} {\bibinfo  {journal} {Phys. Rev. E}\ }\textbf {\bibinfo {volume}
  {99}},\ \bibinfo {pages} {062126} (\bibinfo {year} {2019})}\BibitemShut
  {NoStop}%
\bibitem [{\citenamefont {Dechant}\ and\ \citenamefont
  {Sasa}(2020)}]{dechant2020fluctuation}%
  \BibitemOpen
  \bibfield  {author} {\bibinfo {author} {\bibfnamefont {A.}~\bibnamefont
  {Dechant}}\ and\ \bibinfo {author} {\bibfnamefont {S.-i.}\ \bibnamefont
  {Sasa}},\ }\href@noop {} {\bibfield  {journal} {\bibinfo  {journal}
  {Proceedings of the National Academy of Sciences}\ }\textbf {\bibinfo
  {volume} {117}},\ \bibinfo {pages} {6430} (\bibinfo {year}
  {2020})}\BibitemShut {NoStop}%
\bibitem [{\citenamefont {Hartich}\ and\ \citenamefont
  {Godec}(2021)}]{Hartich2021}%
  \BibitemOpen
  \bibfield  {author} {\bibinfo {author} {\bibfnamefont {D.}~\bibnamefont
  {Hartich}}\ and\ \bibinfo {author} {\bibfnamefont {A.~c.~v.}\ \bibnamefont
  {Godec}},\ }\href {\doibase 10.1103/PhysRevLett.127.080601} {\bibfield
  {journal} {\bibinfo  {journal} {Phys. Rev. Lett.}\ }\textbf {\bibinfo
  {volume} {127}},\ \bibinfo {pages} {080601} (\bibinfo {year}
  {2021})}\BibitemShut {NoStop}%
\bibitem [{\citenamefont {Doerries}\ \emph {et~al.}(2021)\citenamefont
  {Doerries}, \citenamefont {Loos},\ and\ \citenamefont {Klapp}}]{Loos2021}%
  \BibitemOpen
  \bibfield  {author} {\bibinfo {author} {\bibfnamefont {T.~J.}\ \bibnamefont
  {Doerries}}, \bibinfo {author} {\bibfnamefont {S.~A.~M.}\ \bibnamefont
  {Loos}}, \ and\ \bibinfo {author} {\bibfnamefont {S.~H.~L.}\ \bibnamefont
  {Klapp}},\ }\href {\doibase 10.1088/1742-5468/abdead} {\bibfield  {journal}
  {\bibinfo  {journal} {J. Stat. Mech.}\ }\textbf {\bibinfo {volume} {2021}},\
  \bibinfo {pages} {033202} (\bibinfo {year} {2021})}\BibitemShut {NoStop}%
\bibitem [{\citenamefont {Franosch}\ \emph {et~al.}(2011)\citenamefont
  {Franosch}, \citenamefont {Grimm}, \citenamefont {Belushkin}, \citenamefont
  {Mor}, \citenamefont {Foffi}, \citenamefont {Forr{\'o}},\ and\ \citenamefont
  {Jeney}}]{Franosch2011}%
  \BibitemOpen
  \bibfield  {author} {\bibinfo {author} {\bibfnamefont {T.}~\bibnamefont
  {Franosch}}, \bibinfo {author} {\bibfnamefont {M.}~\bibnamefont {Grimm}},
  \bibinfo {author} {\bibfnamefont {M.}~\bibnamefont {Belushkin}}, \bibinfo
  {author} {\bibfnamefont {F.~M.}\ \bibnamefont {Mor}}, \bibinfo {author}
  {\bibfnamefont {G.}~\bibnamefont {Foffi}}, \bibinfo {author} {\bibfnamefont
  {L.}~\bibnamefont {Forr{\'o}}}, \ and\ \bibinfo {author} {\bibfnamefont
  {S.}~\bibnamefont {Jeney}},\ }\href {\doibase 10.1038/nature10498} {\bibfield
   {journal} {\bibinfo  {journal} {Nature}\ }\textbf {\bibinfo {volume}
  {478}},\ \bibinfo {pages} {85} (\bibinfo {year} {2011})}\BibitemShut
  {NoStop}%
\bibitem [{\citenamefont {Zamponi}\ \emph {et~al.}(2005)\citenamefont
  {Zamponi}, \citenamefont {Bonetto}, \citenamefont {Cugliandolo},\ and\
  \citenamefont {Kurchan}}]{zamponi2005fluctuation}%
  \BibitemOpen
  \bibfield  {author} {\bibinfo {author} {\bibfnamefont {F.}~\bibnamefont
  {Zamponi}}, \bibinfo {author} {\bibfnamefont {F.}~\bibnamefont {Bonetto}},
  \bibinfo {author} {\bibfnamefont {L.~F.}\ \bibnamefont {Cugliandolo}}, \ and\
  \bibinfo {author} {\bibfnamefont {J.}~\bibnamefont {Kurchan}},\ }\href@noop
  {} {\bibfield  {journal} {\bibinfo  {journal} {Journal of Statistical
  Mechanics: Theory and Experiment}\ }\textbf {\bibinfo {volume} {2005}},\
  \bibinfo {pages} {P09013} (\bibinfo {year} {2005})}\BibitemShut {NoStop}%
\bibitem [{\citenamefont {Puglisi}\ and\ \citenamefont
  {Villamaina}(2009{\natexlab{a}})}]{puglisi2009irreversible}%
  \BibitemOpen
  \bibfield  {author} {\bibinfo {author} {\bibfnamefont {A.}~\bibnamefont
  {Puglisi}}\ and\ \bibinfo {author} {\bibfnamefont {D.}~\bibnamefont
  {Villamaina}},\ }\href@noop {} {\bibfield  {journal} {\bibinfo  {journal}
  {EPL (Europhysics Letters)}\ }\textbf {\bibinfo {volume} {88}},\ \bibinfo
  {pages} {30004} (\bibinfo {year} {2009}{\natexlab{a}})}\BibitemShut {NoStop}%
\bibitem [{\citenamefont {Crisanti}\ \emph {et~al.}(2012)\citenamefont
  {Crisanti}, \citenamefont {Puglisi},\ and\ \citenamefont
  {Villamaina}}]{crisanti2012nonequilibrium}%
  \BibitemOpen
  \bibfield  {author} {\bibinfo {author} {\bibfnamefont {A.}~\bibnamefont
  {Crisanti}}, \bibinfo {author} {\bibfnamefont {A.}~\bibnamefont {Puglisi}}, \
  and\ \bibinfo {author} {\bibfnamefont {D.}~\bibnamefont {Villamaina}},\
  }\href@noop {} {\bibfield  {journal} {\bibinfo  {journal} {Physical Review
  E}\ }\textbf {\bibinfo {volume} {85}},\ \bibinfo {pages} {061127} (\bibinfo
  {year} {2012})}\BibitemShut {NoStop}%
\bibitem [{\citenamefont {Puglisi}(2014)}]{puglisi2014transport}%
  \BibitemOpen
  \bibfield  {author} {\bibinfo {author} {\bibfnamefont {A.}~\bibnamefont
  {Puglisi}},\ }\href@noop {} {\emph {\bibinfo {title} {Transport and
  fluctuations in granular fluids: From Boltzmann equation to hydrodynamics,
  diffusion and motor effects}}}\ (\bibinfo  {publisher} {Springer},\ \bibinfo
  {year} {2014})\BibitemShut {NoStop}%
\bibitem [{\citenamefont {Rizkallah}\ \emph {et~al.}(2022)\citenamefont
  {Rizkallah}, \citenamefont {Sarracino}, \citenamefont {B{\'e}nichou},\ and\
  \citenamefont {Illien}}]{rizkallah2022microscopic}%
  \BibitemOpen
  \bibfield  {author} {\bibinfo {author} {\bibfnamefont {P.}~\bibnamefont
  {Rizkallah}}, \bibinfo {author} {\bibfnamefont {A.}~\bibnamefont
  {Sarracino}}, \bibinfo {author} {\bibfnamefont {O.}~\bibnamefont
  {B{\'e}nichou}}, \ and\ \bibinfo {author} {\bibfnamefont {P.}~\bibnamefont
  {Illien}},\ }\href@noop {} {\bibfield  {journal} {\bibinfo  {journal}
  {Physical Review Letters}\ }\textbf {\bibinfo {volume} {128}},\ \bibinfo
  {pages} {038001} (\bibinfo {year} {2022})}\BibitemShut {NoStop}%
\bibitem [{\citenamefont {Fischer}\ \emph {et~al.}(2020)\citenamefont
  {Fischer}, \citenamefont {Chun},\ and\ \citenamefont
  {Seifert}}]{Fischer2020}%
  \BibitemOpen
  \bibfield  {author} {\bibinfo {author} {\bibfnamefont {L.~P.}\ \bibnamefont
  {Fischer}}, \bibinfo {author} {\bibfnamefont {H.-M.}\ \bibnamefont {Chun}}, \
  and\ \bibinfo {author} {\bibfnamefont {U.}~\bibnamefont {Seifert}},\ }\href
  {\doibase 10.1103/PhysRevE.102.012120} {\bibfield  {journal} {\bibinfo
  {journal} {Phys. Rev. E}\ }\textbf {\bibinfo {volume} {102}},\ \bibinfo
  {pages} {012120} (\bibinfo {year} {2020})}\BibitemShut {NoStop}%
\bibitem [{\citenamefont {Lee}\ \emph {et~al.}(2021)\citenamefont {Lee},
  \citenamefont {Park},\ and\ \citenamefont {Park}}]{Lee2021}%
  \BibitemOpen
  \bibfield  {author} {\bibinfo {author} {\bibfnamefont {J.~S.}\ \bibnamefont
  {Lee}}, \bibinfo {author} {\bibfnamefont {J.-M.}\ \bibnamefont {Park}}, \
  and\ \bibinfo {author} {\bibfnamefont {H.}~\bibnamefont {Park}},\ }\href
  {\doibase 10.1103/PhysRevE.104.L052102} {\bibfield  {journal} {\bibinfo
  {journal} {Phys. Rev. E}\ }\textbf {\bibinfo {volume} {104}},\ \bibinfo
  {pages} {L052102} (\bibinfo {year} {2021})}\BibitemShut {NoStop}%
\bibitem [{\citenamefont {Dechant}(2022)}]{Dechant2022}%
  \BibitemOpen
  \bibfield  {author} {\bibinfo {author} {\bibfnamefont {A.}~\bibnamefont
  {Dechant}},\ }\href@noop {} {\bibfield  {journal} {\bibinfo  {journal} {arXiv
  preprint arXiv:2202.10696}\ } (\bibinfo {year} {2022})}\BibitemShut {NoStop}%
\bibitem [{\citenamefont {Di~Terlizzi}\ and\ \citenamefont
  {Baiesi}(2020)}]{di2020thermodynamic}%
  \BibitemOpen
  \bibfield  {author} {\bibinfo {author} {\bibfnamefont {I.}~\bibnamefont
  {Di~Terlizzi}}\ and\ \bibinfo {author} {\bibfnamefont {M.}~\bibnamefont
  {Baiesi}},\ }\href@noop {} {\bibfield  {journal} {\bibinfo  {journal}
  {Journal of Physics A: Mathematical and Theoretical}\ }\textbf {\bibinfo
  {volume} {53}},\ \bibinfo {pages} {474002} (\bibinfo {year}
  {2020})}\BibitemShut {NoStop}%
\bibitem [{\citenamefont {Sarracino}\ \emph {et~al.}(2010)\citenamefont
  {Sarracino}, \citenamefont {Villamaina}, \citenamefont {Gradenigo},\ and\
  \citenamefont {Puglisi}}]{sarracino2010irreversible}%
  \BibitemOpen
  \bibfield  {author} {\bibinfo {author} {\bibfnamefont {A.}~\bibnamefont
  {Sarracino}}, \bibinfo {author} {\bibfnamefont {D.}~\bibnamefont
  {Villamaina}}, \bibinfo {author} {\bibfnamefont {G.}~\bibnamefont
  {Gradenigo}}, \ and\ \bibinfo {author} {\bibfnamefont {A.}~\bibnamefont
  {Puglisi}},\ }\href@noop {} {\bibfield  {journal} {\bibinfo  {journal} {EPL
  (Europhysics Letters)}\ }\textbf {\bibinfo {volume} {92}},\ \bibinfo {pages}
  {34001} (\bibinfo {year} {2010})}\BibitemShut {NoStop}%
\bibitem [{\citenamefont {Cortes}\ \emph {et~al.}(1985)\citenamefont {Cortes},
  \citenamefont {West},\ and\ \citenamefont
  {Lindenberg}}]{cortes1985generalized}%
  \BibitemOpen
  \bibfield  {author} {\bibinfo {author} {\bibfnamefont {E.}~\bibnamefont
  {Cortes}}, \bibinfo {author} {\bibfnamefont {B.~J.}\ \bibnamefont {West}}, \
  and\ \bibinfo {author} {\bibfnamefont {K.}~\bibnamefont {Lindenberg}},\
  }\href@noop {} {\bibfield  {journal} {\bibinfo  {journal} {The Journal of
  chemical physics}\ }\textbf {\bibinfo {volume} {82}},\ \bibinfo {pages}
  {2708} (\bibinfo {year} {1985})}\BibitemShut {NoStop}%
\bibitem [{\citenamefont {Munakata}\ and\ \citenamefont
  {Rosinberg}(2013)}]{munakata2013feedback}%
  \BibitemOpen
  \bibfield  {author} {\bibinfo {author} {\bibfnamefont {T.}~\bibnamefont
  {Munakata}}\ and\ \bibinfo {author} {\bibfnamefont {M.}~\bibnamefont
  {Rosinberg}},\ }\href@noop {} {\bibfield  {journal} {\bibinfo  {journal}
  {Journal of Statistical Mechanics: Theory and Experiment}\ }\textbf {\bibinfo
  {volume} {2013}},\ \bibinfo {pages} {P06014} (\bibinfo {year}
  {2013})}\BibitemShut {NoStop}%
\bibitem [{\citenamefont {Munakata}\ and\ \citenamefont
  {Rosinberg}(2014)}]{munakata2014entropy}%
  \BibitemOpen
  \bibfield  {author} {\bibinfo {author} {\bibfnamefont {T.}~\bibnamefont
  {Munakata}}\ and\ \bibinfo {author} {\bibfnamefont {M.}~\bibnamefont
  {Rosinberg}},\ }\href@noop {} {\bibfield  {journal} {\bibinfo  {journal}
  {Physical review letters}\ }\textbf {\bibinfo {volume} {112}},\ \bibinfo
  {pages} {180601} (\bibinfo {year} {2014})}\BibitemShut {NoStop}%
\bibitem [{\citenamefont {Costanzo}\ \emph {et~al.}(2021)\citenamefont
  {Costanzo}, \citenamefont {Lo~Schiavo}, \citenamefont {Sarracino},\ and\
  \citenamefont {Vitelli}}]{costanzo2021stochastic}%
  \BibitemOpen
  \bibfield  {author} {\bibinfo {author} {\bibfnamefont {L.}~\bibnamefont
  {Costanzo}}, \bibinfo {author} {\bibfnamefont {A.}~\bibnamefont
  {Lo~Schiavo}}, \bibinfo {author} {\bibfnamefont {A.}~\bibnamefont
  {Sarracino}}, \ and\ \bibinfo {author} {\bibfnamefont {M.}~\bibnamefont
  {Vitelli}},\ }\href@noop {} {\bibfield  {journal} {\bibinfo  {journal}
  {Entropy}\ }\textbf {\bibinfo {volume} {23}},\ \bibinfo {pages} {677}
  (\bibinfo {year} {2021})}\BibitemShut {NoStop}%
\bibitem [{\citenamefont {Costanzo}\ \emph {et~al.}(2022)\citenamefont
  {Costanzo}, \citenamefont {Lo~Schiavo}, \citenamefont {Sarracino},\ and\
  \citenamefont {Vitelli}}]{Costanzo2022}%
  \BibitemOpen
  \bibfield  {author} {\bibinfo {author} {\bibfnamefont {L.}~\bibnamefont
  {Costanzo}}, \bibinfo {author} {\bibfnamefont {A.}~\bibnamefont
  {Lo~Schiavo}}, \bibinfo {author} {\bibfnamefont {A.}~\bibnamefont
  {Sarracino}}, \ and\ \bibinfo {author} {\bibfnamefont {M.}~\bibnamefont
  {Vitelli}},\ }\href {\doibase 10.3390/e24091222} {\bibfield  {journal}
  {\bibinfo  {journal} {Entropy}\ }\textbf {\bibinfo {volume} {24}} (\bibinfo
  {year} {2022}),\ 10.3390/e24091222}\BibitemShut {NoStop}%
\bibitem [{\citenamefont {Min}\ \emph {et~al.}(2005)\citenamefont {Min},
  \citenamefont {Luo}, \citenamefont {Cherayil}, \citenamefont {Kou},\ and\
  \citenamefont {Xie}}]{min2005observation}%
  \BibitemOpen
  \bibfield  {author} {\bibinfo {author} {\bibfnamefont {W.}~\bibnamefont
  {Min}}, \bibinfo {author} {\bibfnamefont {G.}~\bibnamefont {Luo}}, \bibinfo
  {author} {\bibfnamefont {B.~J.}\ \bibnamefont {Cherayil}}, \bibinfo {author}
  {\bibfnamefont {S.}~\bibnamefont {Kou}}, \ and\ \bibinfo {author}
  {\bibfnamefont {X.~S.}\ \bibnamefont {Xie}},\ }\href@noop {} {\bibfield
  {journal} {\bibinfo  {journal} {Physical review letters}\ }\textbf {\bibinfo
  {volume} {94}},\ \bibinfo {pages} {198302} (\bibinfo {year}
  {2005})}\BibitemShut {NoStop}%
\bibitem [{\citenamefont {Berne}\ \emph {et~al.}(1966)\citenamefont {Berne},
  \citenamefont {Boon},\ and\ \citenamefont {Rice}}]{Berne66}%
  \BibitemOpen
  \bibfield  {author} {\bibinfo {author} {\bibfnamefont {B.~J.}\ \bibnamefont
  {Berne}}, \bibinfo {author} {\bibfnamefont {J.~P.}\ \bibnamefont {Boon}}, \
  and\ \bibinfo {author} {\bibfnamefont {S.~A.}\ \bibnamefont {Rice}},\ }\href
  {\doibase 10.1063/1.1727719} {\bibfield  {journal} {\bibinfo  {journal} {The
  Journal of Chemical Physics}\ }\textbf {\bibinfo {volume} {45}},\ \bibinfo
  {pages} {1086} (\bibinfo {year} {1966})},\ \Eprint
  {http://arxiv.org/abs/https://doi.org/10.1063/1.1727719}
  {https://doi.org/10.1063/1.1727719} \BibitemShut {NoStop}%
\bibitem [{\citenamefont {Gardiner}(2009)}]{Gardiner}%
  \BibitemOpen
  \bibfield  {author} {\bibinfo {author} {\bibfnamefont {C.}~\bibnamefont
  {Gardiner}},\ }\href@noop {} {\emph {\bibinfo {title} {Stochastic Methods}}}\
  (\bibinfo  {publisher} {Springer-Verlag},\ \bibinfo {address} {Berlin},\
  \bibinfo {year} {2009})\BibitemShut {NoStop}%
\bibitem [{\citenamefont {Lebowitz}\ and\ \citenamefont
  {Spohn}(1999)}]{lebowitz1999gallavotti}%
  \BibitemOpen
  \bibfield  {author} {\bibinfo {author} {\bibfnamefont {J.~L.}\ \bibnamefont
  {Lebowitz}}\ and\ \bibinfo {author} {\bibfnamefont {H.}~\bibnamefont
  {Spohn}},\ }\href@noop {} {\bibfield  {journal} {\bibinfo  {journal} {Journal
  of Statistical Physics}\ }\textbf {\bibinfo {volume} {95}},\ \bibinfo {pages}
  {333} (\bibinfo {year} {1999})}\BibitemShut {NoStop}%
\bibitem [{Note1()}]{Note1}%
  \BibitemOpen
  \bibinfo {note} {Note that, due to the linearity, the covariances $\protect
  \hat {\sigma }_{0i}$ do not depend upon $F_{\protect \text
  {ext}}$}\BibitemShut {NoStop}%
\bibitem [{\citenamefont {Wanicharpichat}(2016)}]{Wani2016}%
  \BibitemOpen
  \bibfield  {author} {\bibinfo {author} {\bibfnamefont {W.}~\bibnamefont
  {Wanicharpichat}},\ }\href {\doibase 10.12732/ijpam.v108i4.21} {\ \textbf
  {\bibinfo {volume} {108}} (\bibinfo {year} {2016}),\
  10.12732/ijpam.v108i4.21}\BibitemShut {NoStop}%
\bibitem [{\citenamefont {Cover}(1999)}]{cover1999elements}%
  \BibitemOpen
  \bibfield  {author} {\bibinfo {author} {\bibfnamefont {T.~M.}\ \bibnamefont
  {Cover}},\ }\href@noop {} {\emph {\bibinfo {title} {Elements of information
  theory}}}\ (\bibinfo  {publisher} {John Wiley \& Sons},\ \bibinfo {year}
  {1999})\BibitemShut {NoStop}%
\bibitem [{\citenamefont {Plati}\ and\ \citenamefont
  {Puglisi}(2022)}]{plati2022friction}%
  \BibitemOpen
  \bibfield  {author} {\bibinfo {author} {\bibfnamefont {A.}~\bibnamefont
  {Plati}}\ and\ \bibinfo {author} {\bibfnamefont {A.}~\bibnamefont
  {Puglisi}},\ }\href@noop {} {\bibfield  {journal} {\bibinfo  {journal}
  {Physical review letters}\ }\textbf {\bibinfo {volume} {128}},\ \bibinfo
  {pages} {208001} (\bibinfo {year} {2022})}\BibitemShut {NoStop}%
\bibitem [{\citenamefont {Weiss}(1975)}]{weiss1975time}%
  \BibitemOpen
  \bibfield  {author} {\bibinfo {author} {\bibfnamefont {G.}~\bibnamefont
  {Weiss}},\ }\href@noop {} {\bibfield  {journal} {\bibinfo  {journal} {Journal
  of Applied Probability}\ }\textbf {\bibinfo {volume} {12}},\ \bibinfo {pages}
  {831} (\bibinfo {year} {1975})}\BibitemShut {NoStop}%
\bibitem [{\citenamefont {Lucente}\ \emph {et~al.}(2022)\citenamefont
  {Lucente}, \citenamefont {Baldassarri}, \citenamefont {Puglisi},
  \citenamefont {Vulpiani},\ and\ \citenamefont
  {Viale}}]{lucente2022inference}%
  \BibitemOpen
  \bibfield  {author} {\bibinfo {author} {\bibfnamefont {D.}~\bibnamefont
  {Lucente}}, \bibinfo {author} {\bibfnamefont {A.}~\bibnamefont
  {Baldassarri}}, \bibinfo {author} {\bibfnamefont {A.}~\bibnamefont
  {Puglisi}}, \bibinfo {author} {\bibfnamefont {A.}~\bibnamefont {Vulpiani}}, \
  and\ \bibinfo {author} {\bibfnamefont {M.}~\bibnamefont {Viale}},\ }\href
  {\doibase 10.1103/PhysRevResearch.4.043103} {\bibfield  {journal} {\bibinfo
  {journal} {Phys. Rev. Research}\ }\textbf {\bibinfo {volume} {4}},\ \bibinfo
  {pages} {043103} (\bibinfo {year} {2022})}\BibitemShut {NoStop}%
\bibitem [{\citenamefont {Pietzonka}\ \emph {et~al.}(2017)\citenamefont
  {Pietzonka}, \citenamefont {Ritort},\ and\ \citenamefont
  {Seifert}}]{pietzonka2017finite}%
  \BibitemOpen
  \bibfield  {author} {\bibinfo {author} {\bibfnamefont {P.}~\bibnamefont
  {Pietzonka}}, \bibinfo {author} {\bibfnamefont {F.}~\bibnamefont {Ritort}}, \
  and\ \bibinfo {author} {\bibfnamefont {U.}~\bibnamefont {Seifert}},\
  }\href@noop {} {\bibfield  {journal} {\bibinfo  {journal} {Physical Review
  E}\ }\textbf {\bibinfo {volume} {96}},\ \bibinfo {pages} {012101} (\bibinfo
  {year} {2017})}\BibitemShut {NoStop}%
\bibitem [{\citenamefont {Manikandan}\ \emph {et~al.}(2020)\citenamefont
  {Manikandan}, \citenamefont {Gupta},\ and\ \citenamefont
  {Krishnamurthy}}]{manikandan2020inferring}%
  \BibitemOpen
  \bibfield  {author} {\bibinfo {author} {\bibfnamefont {S.~K.}\ \bibnamefont
  {Manikandan}}, \bibinfo {author} {\bibfnamefont {D.}~\bibnamefont {Gupta}}, \
  and\ \bibinfo {author} {\bibfnamefont {S.}~\bibnamefont {Krishnamurthy}},\
  }\href@noop {} {\bibfield  {journal} {\bibinfo  {journal} {Physical review
  letters}\ }\textbf {\bibinfo {volume} {124}},\ \bibinfo {pages} {120603}
  (\bibinfo {year} {2020})}\BibitemShut {NoStop}%
\bibitem [{\citenamefont {Brandner}\ \emph {et~al.}(2015)\citenamefont
  {Brandner}, \citenamefont {Saito},\ and\ \citenamefont
  {Seifert}}]{brandner2015thermodynamics}%
  \BibitemOpen
  \bibfield  {author} {\bibinfo {author} {\bibfnamefont {K.}~\bibnamefont
  {Brandner}}, \bibinfo {author} {\bibfnamefont {K.}~\bibnamefont {Saito}}, \
  and\ \bibinfo {author} {\bibfnamefont {U.}~\bibnamefont {Seifert}},\
  }\href@noop {} {\bibfield  {journal} {\bibinfo  {journal} {Physical review
  X}\ }\textbf {\bibinfo {volume} {5}},\ \bibinfo {pages} {031019} (\bibinfo
  {year} {2015})}\BibitemShut {NoStop}%
\bibitem [{\citenamefont {Puglisi}\ and\ \citenamefont
  {Villamaina}(2009{\natexlab{b}})}]{Puglisi2009}%
  \BibitemOpen
  \bibfield  {author} {\bibinfo {author} {\bibfnamefont {A.}~\bibnamefont
  {Puglisi}}\ and\ \bibinfo {author} {\bibfnamefont {D.}~\bibnamefont
  {Villamaina}},\ }\href {\doibase 10.1209/0295-5075/88/30004} {\bibfield
  {journal} {\bibinfo  {journal} {{EPL} (Europhysics Letters)}\ }\textbf
  {\bibinfo {volume} {88}},\ \bibinfo {pages} {30004} (\bibinfo {year}
  {2009}{\natexlab{b}})}\BibitemShut {NoStop}%
\bibitem [{\citenamefont {Bochud}\ and\ \citenamefont
  {Challet}(2007)}]{Bochud2007}%
  \BibitemOpen
  \bibfield  {author} {\bibinfo {author} {\bibfnamefont {T.}~\bibnamefont
  {Bochud}}\ and\ \bibinfo {author} {\bibfnamefont {D.}~\bibnamefont
  {Challet}},\ }\href {\doibase 10.1080/14697680701278291} {\bibfield
  {journal} {\bibinfo  {journal} {Quantitative Finance}\ }\textbf {\bibinfo
  {volume} {7}},\ \bibinfo {pages} {585} (\bibinfo {year} {2007})},\ \Eprint
  {http://arxiv.org/abs/https://doi.org/10.1080/14697680701278291}
  {https://doi.org/10.1080/14697680701278291} \BibitemShut {NoStop}%
\end{thebibliography}%

\end{document}